\documentclass[preprintnumbers, floatfix, letterpaper, twocolumn,aps,prd,epsfig,nofootinbib,natbib,longbibliography]{revtex4-1}

\usepackage{graphicx}
\usepackage{epstopdf}
\usepackage{latexsym}
\usepackage{amssymb}
\usepackage{amsmath}
\usepackage{color}
\usepackage{mathrsfs}
\usepackage{xparse}
\usepackage{bbding}
\usepackage{pifont}
\usepackage{comment}
\usepackage{ulem}
\usepackage{float}
\usepackage[inline]{enumitem}
\usepackage{soul}
\usepackage{scalerel}
\usepackage[caption=false]{subfig}

\let\Right\right
\let\Left\left
\makeatletter
\def\right#1{\Right#1\@ifnextchar){\!\right}{}}
\def\left#1{\Left#1\@ifnextchar({\!\left}{}}
\makeatother

\usepackage[
            pdfstartview=FitH,
            bookmarksnumbered=true,
            bookmarksopen=true,
            colorlinks,
            linkcolor=blue,
            anchorcolor=green,
            citecolor=blue
            ]{hyperref}
\begin{document}

  \renewcommand\arraystretch{2}
 \newcommand{\bq}{\begin{equation}}
 \newcommand{\eq}{\end{equation}}
 \newcommand{\bqn}{\begin{eqnarray}}
 \newcommand{\eqn}{\end{eqnarray}}
 \newcommand{\nb}{\nonumber}
 \newcommand{\lb}{\label}
 
\newcommand{\La}{\Lambda}
\newcommand{\va}{\scriptscriptstyle}
\newcommand{\be}{\nopagebreak[3]\begin{equation}}
\newcommand{\ee}{\end{equation}}

\newcommand{\ba}{\nopagebreak[3]\begin{eqnarray}}
\newcommand{\ea}{\end{eqnarray}}

\newcommand{\la}{\label}
\newcommand{\n}{\nonumber}
\newcommand{\su}{\mathfrak{su}}
\newcommand{\SU}{\mathrm{SU}}
\newcommand{\U}{\mathrm{U}}
\newcommand{\red}{ }

\newcommand{\R}{\mathbb{R}}

 \newcommand{\cb}{\color{blue}}
    \newcommand{\cc}{\color{cyan}}
        \newcommand{\cm}{\color{magenta}}
\newcommand{\rc}{\rho^{\scriptscriptstyle{\mathrm{I}}}_c}
\newcommand{\rd}{\rho^{\scriptscriptstyle{\mathrm{II}}}_c} 
\NewDocumentCommand{\evalat}{sO{\big}mm}{%
  \IfBooleanTF{#1}
   {\mleft. #3 \mright|_{#4}}
   {#3#2|_{#4}}%
}
\newcommand{\PRL}{Phys. Rev. Lett.}
\newcommand{\PL}{Phys. Lett.}
\newcommand{\PR}{Phys. Rev.}
\newcommand{\CQG}{Class. Quantum Grav.}

\title{Odd-parity perturbations of the wormhole-like  geometries and quasi-normal modes in Einstein-\AE{}ther theory}


\author{Chao Zhang${}^{a, b, c, d}$}
\email{  {chao123@zjut.edu.cn; a30165@rs.tus.ac.jp}}
\author{Anzhong Wang${}^{e}$\footnote{Corresponding author}}
\email{Anzhong$\_$Wang@baylor.edu; Corresponding author}
\author{Tao Zhu${}^{a, b}$}
\email{zhut05@zjut.edu.cn}

\affiliation{
${}^{a}$ Institute for theoretical physics and cosmology, Zhejiang University of Technology, Hangzhou, 310023, China\\
${}^{b}$ United Center for Gravitational Wave Physics (UCGWP), Zhejiang University of Technology, Hangzhou, 310023, China\\
${}^{c}$ College of Information Engineering, Zhejiang University of Technology, Hangzhou, 310023, China \\
${}^{d}$ Department of Physics, Faculty of Science, Tokyo University of Science,
${}^{e}$GCAP-CASPER, Physics Department, Baylor University, Waco, Texas, 76798-7316, USA}

\date{\today}

\begin{abstract}

The Einstein-$\AE$ther theory has drawn a lot of attentions in recent years. As a representative case of gravitational theories that break the Lorentz symmetry, it plays an important role in testing the Lorentz-violating effects and shedding light on the attempts to construct quantum gravity. Since the first detection to the gravitational wave, the event GW150914, a brand new window has been opened to testing the theory of gravity with gravitational wave observations. At the same time, the study of gravitational waves itself also provides us a serendipity of accessing the nature of a theory.  
In this paper, we focus on the odd-parity gravitational perturbations to a background that describes a wormhole-like geometry under the Einstein-$\AE$ther theory. Taking advantage of this set of analytic background solutions, we are able to simplify the Lagrangian and construct a set of coupled single-parameter dependent master equations, from which we solve for the quasi-normal modes that carry the physical information of the emitted gravitational waves. 
Basically, the results reflect a consistency between Einstein-$\AE$ther theory and general relativity. More importantly, as long as the no-ghost condition and the latest observational constraints are concerned, we notice that the resultant quasi-normal mode solutions intimate a kind of dynamical instability. {Thus, the solutions are ruled out based on their stability against small linear perturbations.}  

\end{abstract}


\maketitle
\section{Introduction}

\renewcommand{\theequation}{1.\arabic{equation}} \setcounter{equation}{0}

The detection of the first gravitational wave (GW) from the coalescence of two massive black holes (BHs) by advanced LIGO/Virgo marked the beginning of a new era --- {\it the GW astronomy}  \cite{Ref1}. Following this observation, about 90 GW events have been identified by the LIGO/Virgo/KAGRA (LVK) scientific collaborations (see, e.g., \cite{GWs, GWs19a, GWs19b, GWsO3b}). In the future, more ground- and space-based GW detectors will be constructed  \cite{Moore2015, Gong:2021gvw}, which will enable us to probe signals with a much wider frequency band and larger distances. This triggered the interest in the quasi-normal mode (QNM) of black holes, as GWs emitted in the ringdown phase can be considered as the linear combination of these individual modes \cite{Berti2009, Berti18}. Similarly, attention has been paid in recent years to QNM originating from wormholes \cite{Kono2018, Volkel2018, Kim2018, Franzin2022}.

From the classical point of view, QNMs are eigenmodes of dissipative systems. The information contained in QNMs provides the keys to revealing whether BHs are ubiquitous in our universe, and more important whether general relativity (GR) is the correct theory to describe gravity even in the strong field regime \cite{Berti18b}. Basically, a QNM frequency $\omega$ contains two parts, the real and imaginary parts. Its real part gives the vibration frequency while its imaginary part provides the damping time.  

In GR,  according to the no-hair theorem, an isolated and stationary BH is completely characterized by only three quantities, mass, angular momentum, and electric charge. Astrophysically, we expect BHs to be neutral, so they are uniquely described by the Kerr solution. Then, the QNM frequencies and damping times will depend only on the mass and angular momentum of the finally formed BH. Clearly, to extract physics from the ringdown phase, at least two QNMs are needed. This will require the signal-to-noise ratio (SNR) to be of the order 100  \cite{Berti18}. Although such high SNRs are not achievable right now, it has been shown that they may be achievable once the advanced LIGO, Virgo, and KAGRA reach their fully designed sensitivities. In any case, it is certain that they will be detected by the ground-based third-generation detectors, such as Cosmic Explorer \cite{CE} and the Einstein Telescope \cite{ET}, as well as the space-based detectors, including LISA \cite{LISA}, TianQin \cite{Liu2020, Shi2019}, Taiji \cite{Taiji2}, and DECIGO \cite{DECIGO}.

QNMs in GR have been studied extensively \cite{Chandra92}, including scalar, vector, and tensor perturbations \cite{Iyer1987}. Such calculations have been extended from the Schwarzschild BH to other more general cases, e.g., the Kerr BH \cite{Det1980, Seidel1990}. In this procedure, several different techniques of computations of QNMs were developed. For instance, the Wentzel-Kramers-Brillouin (WKB) approach \cite{Will1985, Will1987, Konoplya2003, Jerzy2017}, the finite difference method (FDM) \cite{XinLi2020}, the continued fraction method \cite{Leaver1985}, the shooting method \cite{Chandra1975, Doneva2010}, the matrix method \cite{Kai2017}, and so on \cite{Kono2011, Gund1994, Bin2004}. Some of these methods have also been applied to modified theories of gravity \cite{XinLi2020, Oliver2019, KZ07b}. In addition, some special approximations, e.g., the eikonal limit, have also been extensively explored, see, for example, Ref.~\cite{Huan2012} and references therein.
  
This paper focuses on the QNM problem in Einstein-\AE{}ther theory. A set of analytic background solutions describing a throat geometry in the Einstein-\AE{}ther theory will be considered, and the odd-parity perturbations will be investigated. Such studies are well-motivated. In particular, the Einstein-\AE{}ther theory is self-consistent, such as free of ghosts and instability \cite{Jacobson}, and satisfies all the experimental tests carried out so far \cite{OMW18,Tsujikawa21}. Its Cauchy problem is also well-posed \cite{SBP19}, and energy is always positive (as far as the hypersurface-orthogonal \ae{}ther field is concerned) \cite{GJ11}.

In comparison with other modified theories of gravity \cite{CFPS12}, including scalar-tensor theories and their high-order corrections \cite{DL19}, Einstein-\AE{}ther theory has the following distinguishable features: It is a particular vector-tensor theory in which the vector field is always timelike. As a result, it always defines a preferred frame and whereby violates locally the Lorentz invariance (LI).  Despite the facts that LI is the cornerstone of modern physics, and all the experiments carried out so far are consistent with it \cite{Bourgoin21, Bars19, Shao19, Bourgoin17, Flowers17, KR11}, violations of LI have been well motivated and extensively studied in the past several decades, especially from the point of view of quantum gravity \cite{Collin04, Mattingly05, Liberati13, PT14, Wang17}.
In Einstein-\AE{}ther theory there exist three different species of gravitons, spin-0, spin-1, and spin-2, and each of them propagates at different speeds \cite{JM04}.  To avoid  the vacuum gravi-\v{C}erenkov radiation, such as cosmic rays,  each of these three species  must move with a speed that is  at least no less than the speed of light \cite{EMS05}. As a matter of fact, depending on the choice of the free coupling constants of the theory, they can be arbitrarily large, and so far no upper limit of these speeds are known \cite{KR11}.

 With the above remarkable features of the Einstein-\AE{}ther theory, it is interesting and important to find new predictions of the theory for the QNMs mentioned above.  The BH spectroscopy \cite{Berti18} has been extensively studied in the last couple of years in terms of GWs emitted in the ringdown  phase of  binary BHs  (BBHs) (for example, see \cite{BLT21, LCV20, FBPF20, BFPF20, GST19, IGFST19, CDV19, CN20, OC20, Shaik20, Uchi20, BCCK20, Cabero20, Maselli20, Islam20} and references therein), and found that they are all consistent with GR within the error bars allowed by the observations of the 90 GW events \cite{LVK2022}. And here, as mentioned earlier, we shift our insights to the study of wormholes and throat geometries in the Einstein-\AE{}ther theory. 

 The rest of the paper is organized as follows. In Sec.~II we provide a brief introduction to the Einstein-\AE{}ther theory. Fundamental definitions will be given and the background solutions of throat geometries will be discussed. After that, a demonstration of simplifying the odd-parity perturbed Lagrangian under the so-called isotropic coordinate is given in Sec. III. On top of that, we are ready to process the derivations of QNM in Sec. IV. The basic steps of calculations and main results are shown in there. With the results, some concluding remarks are addressed in Sec. V. 
 
We shall adopt the unit system so that $c=G_N=1$, where $c$ is the speed of light and $G_N$ is the Newtonian gravitational constant [It's worth mentioning here that, after this, we still have one degree of freedom (d.o.f) in choosing the unit system for $\{time, length, mass\}$]. We will also work with the {signature} $(-, +,+,+)$.   {All the Greek letters run from 0 to 3.}

\section{Einstein-\AE{}ther theory}
 \renewcommand{\theequation}{2.\arabic{equation}} \setcounter{equation}{0}

In Einstein-\AE{}ther theory ($\ae$-theory), the fundamental variables of the gravitational  sector are \cite{Jacobson},
\bq
\lb{2.0a}
\left(g_{\mu\nu}, u^{\mu}, \lambda\right),
\eq
where $g_{\mu\nu}$ is the four-dimension metric of the spacetime, $u^{\mu}$ is the \ae{}ther field, and $\lambda$ is a Lagrangian multiplier, which guarantees that the \ae{}ther field  is always timelike and unity. Then, the general action of the theory is given by,
\bq
\lb{2.0}
S = S_{\ae} + S_{m},
\eq
where $S_{m}$ denotes the action of matter,  and $S_{\ae}$  the gravitational action of the $\ae$-theory, given, respectively, by
\bqn
\lb{2.1}
S_{\ae} &=& \frac{1}{16\pi G_{\ae} }\int{\sqrt{- g} \; d^4x \Big[  {\cal{L}}_{\ae}\left(g_{\mu\nu}, u^{\alpha}, c_i\right)}\nb\\
&& ~~~~~~~~~~~~ + {\cal{L}}_{\lambda}\left(g_{\mu\nu}, u^{\alpha}, \lambda\right)\Big],\nb\\
S_{m} &=& \int{\sqrt{- g} \; d^4x \Big[{\cal{L}}_{m}\left(g_{\mu\nu}, u^{\alpha}; \hat\psi\right)\Big]}.
\eqn
Here $\hat\psi$ collectively denotes the matter fields, and $g$ is the determinant of $g_{\mu\nu}$, and
\bqn
\lb{2.2}
{\cal{L}}_{\lambda}  &\equiv&  \lambda \left(g_{\alpha\beta} u^{\alpha}u^{\beta} + 1\right),\nb\\
{\cal{L}}_{\ae}  &\equiv& R(g_{\mu\nu}) - M^{\alpha\beta}_{~~~~\mu\nu}\left(D_{\alpha}u^{\mu}\right) \left(D_{\beta}u^{\nu}\right),
\eqn
where $D_{\mu}$ denotes the covariant derivative with respect to $g_{\mu\nu}$, $R$ is the Ricci scalar, and  $M^{\alpha\beta}_{~~~~\mu\nu}$ is defined as
\bqn
\lb{2.3}
M^{\alpha\beta}_{~~~~\mu\nu} \equiv c_1 g^{\alpha\beta} g_{\mu\nu} + c_2 \delta^{\alpha}_{\mu}\delta^{\beta}_{\nu} +  c_3 \delta^{\alpha}_{\nu}\delta^{\beta}_{\mu} - c_4 u^{\alpha}u^{\beta} g_{\mu\nu},\nb\\
\eqn
with $\delta_{\mu \nu}$ representing the Kronecker delta. Note that here we assume that matter fields couple not only to $g_{\mu\nu}$ but also to the \ae{}ther field $u^{\mu}$. However, in order to satisfy the severe observational constraints,  such a coupling, in general, is assumed to be absent  \cite{Jacobson}.

The four coupling constants $c_i$'s are all dimensionless, and $G_{\ae} $ is related to the Newtonian gravitational constant $G_{N}$ via the relation \cite{CL04},
\bq
\lb{2.3a}
G_{N} = \frac{G_{\ae}}{1 - \frac{1}{2}c_{14}},
\eq
where $c_{ij} \equiv c_i+c_j$.

\subsection{Field Equations}

The variations of the total action, respectively, with respect to $g_{\mu\nu}$,  $u^{\mu}$   and $\lambda$ yield respectively the field equations \cite{Chao2020b},
\bqn
\lb{2.4a}
R^{\mu\nu} - \frac{1}{2} g^{\mu\nu}R - S^{\mu\nu} &=& 8\pi G_{\ae}  T^{\mu\nu},\\
\lb{2.4b}
\AE_{\mu} &=& 8\pi G_{\ae}  T_{\mu}, \\
\lb{2.4c}
g_{\alpha\beta} u^{\alpha}u^{\beta} &=& -1,
\eqn
where $R^{\mu \nu}$ denotes the Ricci tensor, and
\bqn
\lb{2.5}
S_{\alpha\beta} &\equiv&
D_{\mu}\Big[J^{\mu}_{\;\;\;(\alpha}u_{\beta)} + J_{(\alpha\beta)}u^{\mu}-u_{(\beta}J_{\alpha)}^{\;\;\;\mu}\Big]\nb\\
&& + c_1\Big[\left(D_{\alpha}u_{\mu}\right)\left(D_{\beta}u^{\mu}\right) - \left(D_{\mu}u_{\alpha}\right)\left(D^{\mu}u_{\beta}\right)\Big]\nb\\
&& + c_4 a_{\alpha}a_{\beta}    + \lambda  u_{\alpha}u_{\beta} - \frac{1}{2}  g_{\alpha\beta} J^{\delta}_{\;\;\sigma} D_{\delta}u^{\sigma},\nb\\
\AE_{\mu} & \equiv &
D_{\alpha} J^{\alpha}_{\;\;\;\mu} + c_4 a_{\alpha} D_{\mu}u^{\alpha} + \lambda u_{\mu},\nb\\
T^{\mu\nu} &\equiv&  \frac{2}{\sqrt{-g}}\frac{\delta \left(\sqrt{-g} {\cal{L}}_{m}\right)}{\delta g_{\mu\nu}},\nb\\
T_{\mu} &\equiv& - \frac{1}{\sqrt{-g}}\frac{\delta \left(\sqrt{-g} {\cal{L}}_{m}\right)}{\delta u^{\mu}},
\eqn
with
\begin{equation}
\lb{2.6}
J^{\alpha}_{\;\;\;\mu} \equiv M^{\alpha\beta}_{~~~~\mu\nu}D_{\beta}u^{\nu}\,,\quad
a^{\mu} \equiv u^{\alpha}D_{\alpha}u^{\mu}.
\end{equation}
From Eq.(\ref{2.4b}),  we find that
\bq
\lb{2.7}
\lambda = u_{\beta}D_{\alpha}J^{\alpha\beta} + c_4 a^2 - 8\pi G_{\ae}  T_{\alpha}u^{\alpha},
\eq
where $a^{2}\equiv a_{\lambda}a^{\lambda}$. Notice that, by considering only the vacuum solutions (as what we will do later), the matter fields disappear, leading to the absence of $T^{\mu \nu}$ as well as $T_\mu$ in Eqs. \eqref{2.4a} and \eqref{2.4b}.

It is easy to show that the Minkowski spacetime is a solution of $\ae$-theory, in which the \ae{}ther is aligned along the time direction, ${u}_{\mu} = \delta^{0}_{\mu}$. Then, the linear perturbations around the Minkowski background show that the theory in general possesses three types of excitations, scalar  (spin-0), vector (spin-1), and tensor (spin-2)
modes, with their squared  speeds given by   \cite{JM04}
 \begin{eqnarray}
\label{speeds}
c_S^2 & = & \frac{c_{123}(2-c_{14})}{c_{14}(1-c_{13}) (2+c_{13} + 3c_2)}\,,\nonumber\\
c_V^2 & = & \frac{2c_1 -c_{13} (2c_1-c_{13})}{2c_{14}(1-c_{13})}\,,\nonumber\\
c_T^2 & = & \frac{1}{1-c_{13}},
\end{eqnarray}
respectively. Here $c_{ijk} \equiv c_i+c_j+c_k$.

Requiring that the theory: 1)  be self-consistent, such as free of ghosts; and 2) satisfies all the observational constraints obtained so far, it was found that the parameter space of the theory is considerably restricted. In particular, $c_{14}$, $c_2$ and $c_{13}$  are restricted to \cite{OMW18} \footnote{The recent studies of the neutron binary systems showed that the PPN parameter $\alpha_1$ is further restricted to $|\alpha_1| < 10^{-5}$ \cite{GHBBCYY}, which is an order of magnitude   stronger than the bounds from lunar laser ranging experiments \cite{MWT08}. This will translate  the constraint on $c_{14}$ given by Eq.(\ref{c1234a}) to $0 \lesssim c_{14} \lesssim  \times 2.5 \times 10^{-6}$, as one can see clearly from Eq.(3.12) given  in \cite{OMW18}.},
\bqn
\lb{c1234a}
&& 0 \lesssim c_{14} \lesssim 2.5 \times 10^{-5},\\
\lb{c1234ab}
&&  c_{14} \lesssim c_2 \lesssim  0.095,\\
\lb{c1234ac}
&& \left|c_{13}\right|  \lesssim 10^{-15}.
\eqn
{Taking $c_{13} = 0$,  the stability of the odd-parity perturbations of BHs further requires  $c_4 = 0$   \cite{Tsujikawa21}.}

\subsection{Background Geometry}

From now on, we shall consider solely the vacuum solutions to the field equations [cf., Eqs.\eqref{2.4a}-\eqref{2.4c}]. For later convenience, in this subsection, we shall first adopt the isotropic coordinate, represented by $x^\mu = (t, \rho, \theta, \phi)$ (see, e.g., \S8.2 of \cite{Weinberg1972}). Thus, for the spherically symmetric time-independent case, the background line element and the \ae{}ther field are given by \cite{Jacob2021}
\bqn
\lb{backeqn1}
ds^2 &=& -e^{2 \mu(\rho)} dt^2+e^{2\nu(\rho)} \left(d\rho^2 + \rho^2 d \Omega^2\right), \nb\\
u^\alpha &=& e^{-\mu(\rho)} \delta^\alpha_0,
\eqn
where $d\Omega^2=d\theta^2+\sin^2\theta d \phi^2$.

As pointed in \cite{Jacob2021}, substituting \eqref{backeqn1} into Eq.\eqref{2.4a}, and picking up the $\rho \rho$- \footnote{Notice that, in \cite{Jacob2021} the coordinate ``$\rho$'' was written as ``$r$''.} and $\theta \theta$-components, we obtain
\bqn
\lb{backeqn2a}
0 &=& \frac{1}{2} c_{14} \left(\mu '\right)^2+2 \mu ' \nu '+\left(\nu '\right)^2+\frac{2 \mu '}{r}+\frac{2 \nu '}{r}, \nb\\
0 &=& \left(1-\frac{c_{14}}{2}\right) \left(\mu '\right)^2+\mu ''+\nu ''+\frac{\mu '}{r}+\frac{\nu '}{r},
\eqn
\break
where a prime in the superscript expediently stands for the derivative respective to $\rho$. 

Clearly, a master equation could be easily constructed from Eq.\eqref{backeqn2a}. As shown in \cite{Jacob2021}, that leads to the solutions
\bqn
\lb{backeqn2}
\mu &=&  ({\bar q}+1) \ln \left(\frac{r-r_0}{r+r_0}\right), \nb\\
\nu &=& \ln \left[\left(1-\frac{r_0^2}{r^2}\right) \left(\frac{r-r_0}{r+r_0}\right){}^{-({\bar q}+1)}\right],
\eqn
where $ {\bar q} \equiv \sqrt{1/(1-c_{14}/2)}-1\in [0, 6.25\times 10^{-7})$[cf., Eq.\eqref{c1234a} and footnote \#1] and $r_0 \equiv m/2$ \cite{Weinberg1972, Jacob2021}. Note that, the solutions in \eqref{backeqn2} are characterized by a single parameter $c_{14}$. In addition, they describe a throat geometry. As long as $c_{14} \neq 0$, a marginally trapped throat with a finite non-zero radius always exists, and on one side of it the spacetime is asymptotically flat, while on the other side, the spacetime becomes singular within a finite proper distance from the throat \cite{Jacob2021}.

When moving to the Schwarzschild coordinate, represented by $x^\mu = (t, r, \theta, \phi)$, the line element could be written as
\bqn
\lb{backgab1}
ds^2 &=& -\left(1-\frac{2m}{r}\right)^{{\bar q}+1} dt^2+\left(1-\frac{2m}{r}\right)^{-({\bar q}+1)} dr^2 \nb\\
&& + \left(1-\frac{2m}{r}\right)^{-{\bar q}} r^2 d \Omega^2, 
\eqn
which will immediately go back to that of GR at the ${\bar q} \to 0$ limit \cite{Jacob2021}.

\begin{figure}[htb]
	\includegraphics[width=\columnwidth]{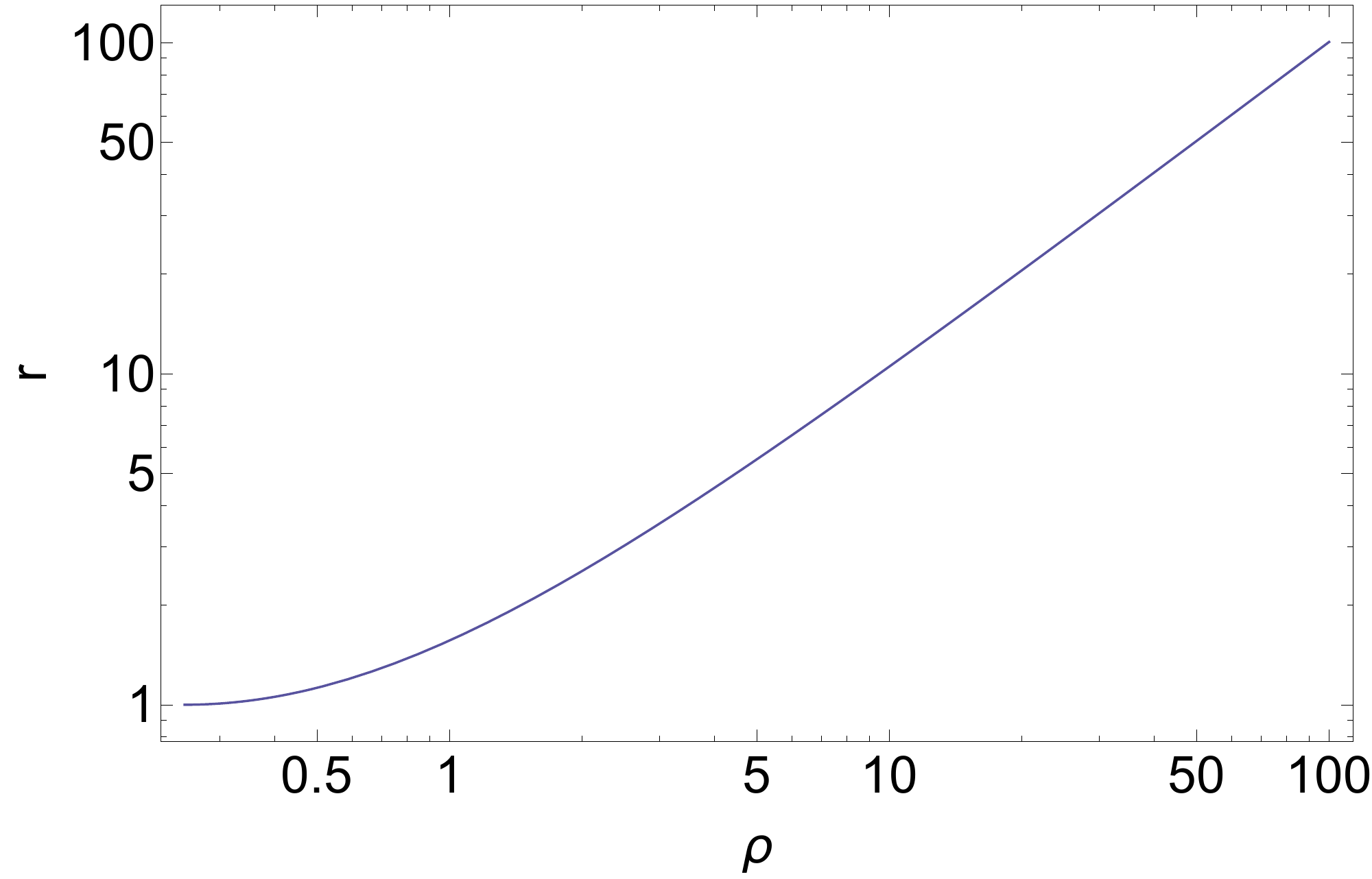} 
\caption{The relation between $r$ and $\rho$, where we have set $r_0=1/4$.} 
	\label{plot2}
\end{figure}

Notice that, a simple relation between $r$ and $\rho$ is given by \cite{Weinberg1972}
\bqn
\lb{randrho}
r &=& \rho \left(1+\frac{r_0}{\rho}\right)^2.
\eqn
To make it more clear, we show the relation between $r$ and $\rho$ in Fig. \ref{plot2}, in which we set $r_0=1/4$. 

\section{The odd-parity perturbations}
 \renewcommand{\theequation}{3.\arabic{equation}} \setcounter{equation}{0}

In this section, we move to the odd-parity perturbations to the background described by Eq.\eqref{backeqn1}. The full metric and \ae{}ther field are given like
\bqn
\lb{gab1}
g_{\mu \nu} &=& {\bar g}_{\mu \nu}+ \epsilon h_{\mu \nu}, \nb\\
u_{\mu} &=& {\bar u}_\mu + \epsilon w_\mu.
 \eqn
Here, the ${\bar g}_{\mu \nu}$ and  ${\bar u}_\mu$ denote the background fields, described by Eq.\eqref{backeqn2}, while $\epsilon$ stands for an infinitesimal constant. 

Firstly, we shall keep working in the isotropic coordinate. By mimicking \cite{Thomp2017}, the perturbation terms could be written as
\begin{widetext}
	\bqn
	\lb{hab}
	h_{\mu \nu} &=&  \sum_{l=0}^{\infty} \sum_{m=-l}^{l}
	\begin{pmatrix}
		0	& 0 &  C_{lm}\csc \theta \partial_\phi & -C_{lm} \sin \theta \partial_\theta\\
		0	& 0 &  J_{lm}\csc \theta \partial_\phi & -J_{lm} \sin \theta \partial_\theta\\
		sym	& sym &  G_{lm}\csc \theta \big(\cot \theta \partial_\phi-\partial_\theta \partial_\phi \big) &  sym \\
		sym	& sym &  \frac{1}{2} G_{lm}\big(\sin \theta \partial^2_\theta - \cos \theta \partial_\theta- \csc \theta \partial^2_\phi\big) &   - G_{lm} \sin \theta \big(\cot \theta \partial_\phi-  \partial_\theta \partial_\phi \big)
	\end{pmatrix}
	Y_{l m}(\theta, \phi), 	~~~
\eqn
\end{widetext}
and
\bqn
\lb{wa}
	w_\mu &=& \sum_{l=0}^{\infty} \sum_{m=-l}^{l}
	\begin{pmatrix}
		0	\\
		0	\\
		a_{lm} \csc \theta  \partial_\phi 	\\
		-a_{lm} \sin \theta  \partial_\theta
	\end{pmatrix}
Y_{l m}(\theta, \phi),
\eqn
where $a_{lm}$, $C_{lm}$, $J_{lm}$ and $G_{lm}$ are functions of $t$ and $\rho$\footnote{
It must not be confused with  the function {$J_{lm}$} introduced  here  and the tensor {$J_{\alpha \beta}$} appearing in Eq.\eqref{2.5}.}, while $Y_{l m}(\theta, \phi)$ denotes the spherical harmonics. Starting from now on, we shall set  $m=0$ in the above expressions so that $\partial_\phi Y_{lm}(\theta, \phi)=0$, as now the background has the spherical symmetry, and the corresponding linear perturbations do not depend on $m$   \cite{Regge57,Thomp2017} \footnote{Notice that, since we are using the isotropic coordinate and following some different conventions, the terms $\{C_{lm}, J_{lm}, G_{lm}\}$ in \eqref{hab} are not necessarily equal to their counterparts given in \cite{Thomp2017}.}.

\subsection{Gauge Transformations and Gauge Fixing}
For later convenience, we here investigate the infinitesimal gauge transformations (Recall that we only consider the odd-parity perturbations and $m=0$)
\bqn
\lb{gauge1}
x^\alpha \to x^{\prime \alpha} =x^\alpha+  \epsilon \xi^{\alpha},
\eqn
where
\bqn
\lb{xi}
\xi^\alpha &=& - \frac{\csc \theta \partial_\theta Y_{lm}(\theta, \phi)}{r^2}  \left(0,0,0,1 \right) e^{-2 \nu} \Lambda,~~~~
\eqn
with $\Lambda$ being a function of $t$ and $\rho$.
Under the transformation of Eq.(\ref{gauge1}), we have  
\bqn
\lb{deltau}
&& \Delta w_\mu \equiv \left(w_\mu\right)_{new}-\left(w_\mu\right)_{old} =- {\cal{L}}_{\xi} {\bar u}_\mu, \nb\\
&& \Delta h_{\mu\nu} \equiv \left(h_{\mu\nu}\right)_{new}-\left(h_{\mu\nu}\right)_{old} =- {\cal{L}}_{\xi} {\bar g}_{\mu\nu},
\eqn
\break
where $\cal{L}$ stands for the Lie derivative \cite{Invb}.  From Eq.(\ref{deltau}) we find
\bqn
\lb{gauge2}
&& \Delta C_{lm} \equiv \left( C_{lm}\right)_{old}-\left( C_{lm}\right)_{new} = {\dot \Lambda}, \nb\\
&& \Delta G_{lm} \equiv \left( G_{lm}\right)_{old}-\left( G_{lm}\right)_{new} = -2 \Lambda, \nb\\
&& \Delta J_{lm} \equiv \left( J_{lm}\right)_{old}-\left( J_{lm}\right)_{new}  = \Lambda' - \frac{2(1+r \nu')}{r} \Lambda, \nb\\
&& \Delta a_{lm} \equiv \left( a_{lm}\right)_{old}-\left( a_{lm}\right)_{new} = 0,
\eqn
where a prime and a dot stand for the derivatives with respect to $\rho$ and $t$, respectively.
 With the above gauge transformations,  we can construct the gauge-invariant (GI) quantities, and due to the presence of the \ae{}ther field,
 three such independent quantities can be constructed, in contrast to the relativistic case, in which only two such quantities can be constructed.
 These three gauge invariants  can be  defined as
\bqn
\lb{gauge4}
{\cal{X}}_{lm} (t, \rho) &\equiv& {C}_{lm} +  \frac{1}{2} {\dot G}_{lm}, \nb\\
{\cal{Y}}_{lm} (t, \rho) &\equiv&  {J}_{lm} +  \frac{1}{2} {G}_{lm}'-\left(\frac{1}{r}+\nu'\right) {G}_{lm}, \nb\\
{\cal{Z}}_{lm} (t, \rho) &\equiv& a_{lm}.
\eqn
Of course, any combination of these quantities is also GI. According to Eq.\eqref{gauge4}, by simply choosing the gauge condition ${G}_{lm}=0$ (This could be referred as the RW gauge \cite{Thomp2017}), ${\cal{X}}_{lm}$ and ${\cal{Y}}_{lm}$ will reduce to $C_{lm}$ and $J_{lm}$, respectively. This will be the gauge condition that we shall adopt in the following of this paper. 
\break

\subsection{Simplified Lagrangian}
To derive the partial differential equations (PDEs) for the perturbations, we need to first simplify the original Lagrangian, adopted in \eqref{2.0}. Up to the 2nd order of $\epsilon$, the total action can be cast in the form 
\bqn
\lb{action1}
{S} &=& {S}_{(0)} + \epsilon {S}_{(1)} + \epsilon^2 { S}_{(2)} + {\cal O}(\epsilon^3).
\eqn
Following \cite{Tsujikawa21}, we substitute  \eqref{gab1} [together with Eqs.\eqref{backeqn1}, \eqref{hab} and \eqref{wa}] into \eqref{2.0}, and then pick up the ${\cal O}(\epsilon^2)$ terms. In addition, from now on, we shall set
\bqn
\lb{c13}
c_{13} &=& 0,
\eqn
since it has been confined to an extremely narrow region [cf., \eqref{c1234ac}]. The resultant 2nd-order action is given by
\bqn
\lb{action2}
 && {\cal S}_{(2)} = \int dt d\rho {\cal L}_{(2)},\\
 \lb{action3}
&& {\cal L}_{(2)} = {\upsilon} {\cal L}_{\text{odd}}, \quad {\upsilon} \equiv {\frac{r^2 e^{\mu+3\nu}}{16 \pi G_{\ae}}}.
 \eqn
 Notice that, the Lagrangian ${\cal L}_{(2)}$ is only a function of $t$ and $\rho$, as $\theta$ and $\phi$ have been integrated out, during which procedure, the features of spherical harmonics have been used intensively (see Appendix A).  

 With the help of integration by parts \cite{Arfken}, plus the background field equations [cf., \eqref{2.4a}-\eqref{2.4c}], the quantity $ {\cal L}_{\text{odd}}$ is simplified to
 \begin{widetext}
 \bqn
\lb{action4}
{\cal L}_{\text{odd}} &=& \alpha_1 {\dot a}_{lm}^2 +  \alpha_2 a_{lm}^{\prime 2}  + \alpha_3 \chi {\dot J}_{lm} + \alpha_4 \chi {C}_{lm}'  + \alpha_5 a_{lm}^2 + \alpha_6 \chi^2 + \alpha_7 J_{lm}^2 + \alpha_8 C_{lm}^2  + \alpha_9 \chi a_{lm} + \alpha_{10} \chi {C}_{lm} \nb\\
 &=& \alpha_1 {\dot a}_{lm}^2 +  \alpha_2 a_{lm}^{\prime 2}  + \alpha_3 \chi {\dot J}_{lm} + \alpha_4 \chi {C}_{lm}' + \alpha_5 a_{lm}^2 - \alpha_6 \chi^2 + \alpha_7 J_{lm}^2 + \alpha_8 C_{lm}^2 + \alpha_9 \chi a_{lm} + \alpha_{10} \chi {C}_{lm} \nb\\
&& +2  \alpha_6 \chi \left[ 2 c_{14} e^{\mu } \mu' a_{lm}-C_{lm}'+2 \left(\nu '+\frac{1}{\rho}\right) C_{lm}+ {\dot J}_{lm} \right],
 \eqn
 \end{widetext}
 where
 \bqn
\lb{chi}
\chi(t, \rho) &\equiv& 2 c_{14} e^{\mu } \mu ' a_{lm}-C_{lm}'+2 \left(\nu '+\frac{1}{\rho}\right) C_{lm}+ {\dot J}_{lm}, \nb\\
 \eqn
which could be matched to its counterpart in \cite{Tsujikawa21}. Recall that $C_{lm}$, $J_{lm}$ and $a_{lm}$ (as well as their combination $\chi$) are GI under the chosen gauge condition, viz., $G_{lm}=0$ [cf., Eq.\eqref{gauge4}]. Note that, before stimulating any confusion, we shall omit the subscript $lm$ for $\chi$. The coefficients $\alpha_i$'s are functions of $\rho$, $L$, $c_{14}$, $c_1$, etc. We abbreviate their full expressions since they are only for the narration of intermediate steps. 

In \eqref{action4}, the apparent d.o.f for the odd-parity perturbations is 3, spanned by $\{J_{lm}, C_{lm}, a_{lm}\}$. However, the true d.o.f is 2 \cite{Tsujikawa21}. To get rid of that redundant d.o.f, we apply the Euler-Lagrange (E-L) equation \cite{Taylor05} to $ {\cal L}_{(2)}$ with respect to $J_{lm}$ and $C_{lm}$, and obtain 
 \bqn
\lb{JandC}
J_{lm} &=& e^{2 \nu } \rho^2 {\dot \chi} \nb\\
&& \times \Big\{ e^{2 \mu } \Big[ c_{14} \rho^2 \left(\mu '\right)^2-L+2 \rho^2 \left(\nu '\right)^2 \nb\\
&& +4 \rho \mu ' \left(\rho \nu '+1\right)+4 \rho \nu '+2\Big] \Big\}^{-1}, \nb\\
C_{lm} &=& \frac{\rho^2 }{2-L} \chi'+\frac{\rho  \left(\rho \mu '-\rho \nu '-2\right)}{L-2} \chi,
\eqn
where we have defined $L \equiv l(l+1)$. Substitute \eqref{JandC} into the second line of \eqref{action4}, together with some rearrangements, ${\cal L}_{\text{odd}} $ becomes
 \bqn
\lb{Lodd}
{\cal L}_{\text{odd}} &=& \beta_1 {\dot \chi}^2 + \beta_2 {\dot a}_{lm}^2 + \beta_3 {\chi}^{\prime 2} + \beta_4 {a}_{lm}^{\prime 2} \nb\\
&& +  \beta_5 {\chi}^2 +  \beta_6 {a}_{lm}^2  +  \beta_7 \chi {a}_{lm}.
\eqn
The coefficients $\beta_i$'s could be found in Appendix B. Clearly, as promised earlier, the current apparent d.o.f for the 2nd-order Lagrangian is 2, spanned by $\{\chi, a_{lm}\}$.


\section{Quasi-normal modes of the odd-parity perturbations}
 \renewcommand{\theequation}{4.\arabic{equation}} \setcounter{equation}{0}

\subsection{Master Equations} 
Combining \eqref{action3} and \eqref{Lodd}, we obtain the simplified 2nd-order Lagrangian for the $c_{13}=0$ case. Now we are ready to derive the master equation for calculating the QNMs. With this Lagrangian, we apply the E-L equation to it with respect to $\chi$ and $a_{lm}$. That leads to a set of coupled PDEs
 \bqn
\lb{master1}
0 &=& - {\ddot \chi} + \gamma_{11} \chi'' + \gamma_{12} \chi'+ \gamma_{13} \chi + \gamma_{14} a_{lm}, \nb\\
0 &=& - {\ddot a}_{lm} + \gamma_{21} a_{lm}'' + \gamma_{22} a_{lm}'+ \gamma_{23} a_{lm} + \gamma_{24} \chi, ~~~~
\eqn
where a prime in the superscript expediently denotes the derivative with respect to $\rho$. The expressions of $\gamma_{ij}$'s could be found in Appendix C.

In the next step, we shall apply the analytic solutions \eqref{backeqn2}, and transfer our PDEs from the isotropic coordinate to the Schwarzschild coordinate $(t, r, \theta, \phi)$. In addition, taking advantage of the residual d.o.f for choosing the unit system, we further set $m=1/2$ (so that the unit system of $\{time, length, mass\}$ is totally fixed). By doing so, the PDEs are translated to
\begin{widetext}
 \bqn
\lb{master2}
0 &=& - {\ddot \chi} + \left(\frac{r-1}{r}\right)^{\bar{q}+2} \chi'' + \frac{2 }{r} \left(\frac{r-1}{r}\right)^{\bar{q}+\frac{3}{2}} \chi' \nb\\
&& + \frac{1}{4} (r-1)^{\bar{q}} r^{-\bar{q}-4} \Big[ (8 r-4) \left(\bar{q}+2\right)-\left(\bar{q}+2\right)^2+4 r (L (-r)+L+r-1)-4 \sqrt{(r-1) r^3}+2 \sqrt{(r-1) r}-3 \Big] \chi \nb\\
&&   -\frac{4 (L-2) \left[-2 r+2 \sqrt{(r-1) r}+1\right] \left[\left(\bar{q}+2\right)^2-4\right] }{[(r-1) r]^{3/2} \left(\bar{q}+2\right)} \left(\frac{r-1}{r}\right)^{\frac{5}{4} \left(\bar{q}+2\right)} a_{lm}, \nb\\
0 &=& - {\ddot a}_{lm} + \frac{c_1 \left(\bar{q}+2\right)^2 }{2 \left[\left(\bar{q}+2\right)^2-4\right]} \left(\frac{r-1}{r}\right)^{\bar{q}+2} a_{lm}'' + \frac{c_1 \left(\bar{q}+2\right)^3 (r-1)^{\bar{q}+1} r^{-\bar{q}-3}}{4 \left[\left(\bar{q}+2\right)^2-4\right]} a_{lm}' \nb\\
&& + \frac{(r-1)^{\bar{q}} r^{-\bar{q}-4} }{32 \left[\left(\bar{q}+2\right)^2-4\right]} a_{lm} \nb\\
&& \times \left\{ \left(\bar{q}+2\right) \left[\left(\bar{q}+2\right) \left(4 \left(c_1-2\right) (2 r-1) \left(\bar{q}+2\right)-3 c_1 \left(\bar{q}+2\right)^2-16 c_1 L (r-1) r+32\right)+64 r-32\right]-128\right\} \nb\\
&& + \frac{1}{16} \left[2 r+2 \sqrt{(r-1) r}-1\right] \left(\bar{q}+2\right) (r-1)^{\frac{3 \bar{q}}{4}} r^{\frac{1}{4} (-3) \left(\bar{q}+4\right)} \chi, ~~~~
\eqn
\end{widetext}
where a prime in the superscript now stands for the derivative with respect to $r$. A dot, as usual, stands for the derivative with respect to $t$. 

{In writing down  Eq.(\ref{master2}), the stability condition of BHs that found in \cite{Tsujikawa21}, viz., $c_4 = 0$, needs to be considered [so that the only theory-dependent coupling constant of Eq.(\ref{master2}) is $c_{1}$ (or equivalently, $\bar q$), since now we have $c_1=c_{14}$]. }
By performing suitable coordinate transformations, the PDEs in \eqref{master2} could be further translated to [the formulas in Appendix D were used in deriving Eqs.\eqref{master3a} and \eqref{master3b}]
 \bqn
\lb{master3a}
Q_1 \Psi_2  &=& \frac{\partial^2 \Psi_1}{\partial x^2} - \left[\frac{\partial^2}{\partial t^2} + V_1(r) \right]\Psi_1,  \\ 
\lb{master3b}
Q_2 \Psi_1  &=&    \frac{\partial^2 \Psi_2}{\partial x^2} - \left[\frac{\partial^2}{\partial t^2} + V_2(r) \right]\Psi_2, ~~
\eqn
where
 \bqn
\lb{q1q2}
\frac{dr}{dx} &=& \left(\frac{r-1}{r}\right)^{\frac{1}{2} \left(\bar{q}+2\right)}, \nb\\
\Psi_1 &\equiv& \frac{1}{-2 r+2 \sqrt{(r-1) r}+1} \left(\frac{r-1}{r}\right)^{\frac{1}{4} \left(-\bar{q}-2\right)} \chi, \nb\\
\Psi_2 &\equiv& a_{lm}, 
\eqn
and
\begin{widetext}
 \bqn
\lb{V1V2}
V_1 &\equiv& \frac{ \left[-24 (r-1) \bar{q}+3 \bar{q}^2+16 (r-1) (L r-3)\right]}{16 r^4} \left(\frac{r-1}{r}\right)^{\bar{q}},\nb\\
V_2 &\equiv&   \Big\{ \left(\bar{q}+2\right) \left[\left(\bar{q}+2\right) \left(4 \left(\bar{q}+2\right) \left(-2 c_1 r+c_1+4 r-2\right)+3 c_1 \left(\bar{q}+2\right)^2+16 c_1 L (r-1) r-32\right)-64 r+32\right] \nb\\
&& ~~~ +128\Big\} \left(\frac{r-1}{r}\right)^{\bar{q}} \frac{ 1}{32 r^4 \bar{q} \left(\bar{q}+4\right)}, \nb\\
Q_1 &\equiv& \frac{4 (L-2) \bar{q} \left(\bar{q}+4\right) (r-1)^{\bar{q}+\frac{1}{2}} r^{-\bar{q}-\frac{7}{2}}}{\bar{q}+2}, \nb\\
Q_2 &\equiv& \frac{1}{16} \left(\bar{q}+2\right) (r-1)^{\bar{q}+\frac{1}{2}} r^{-\bar{q}-\frac{7}{2}}.
\eqn
\end{widetext}
It's worth mentioning here that, at the ${\bar q} \to 0$ limit, Eq.\eqref{master3a} will reduce precisely to that of GR (see, e.g., the Eq.(2.13) of \cite{Kono2011}).


\subsection{Calculate for the QNMs} 

 {As has been shown, taking $c_{13}=c_4=0$ brings us the coupled PDEs, Eqs.\eqref{master3a} and \eqref{master3b}.} 
With the above master equations, we are now at the stage of solving them for QNMs. To deal with this set of coupled PDEs, we shall apply the finite difference method (FDM) \cite{XinLi2020, Habermanb}.  By working with the FDM, we are expecting to solve \eqref{master3a} and \eqref{master3b} for $\Psi_{1,\;2}$. The solved $\Psi_{1,\;2}$ will carry the information of QNM frequency $\omega$ (see, e.g., \cite{Berti2009} for a review). Notice that,  $\Psi_{1,\;2}$ can only reflect the comprehensive effects of all the existing $\omega$'s and it's not trivial to extract individual $\omega$'s from $\Psi_{1,\;2}$.
 

 To perform FDM, we shall basically follow \cite{Habermanb, Kai2016}, and obtain the recursion formula
\begin{widetext}
\bqn
\lb{FDM2}
\Psi_{1,\;2}(t+\delta t, x) &\cong & -\Psi_{1,\;2}(t-\delta t, x) - \left[ 2 \left( \mathfrak{v}_{1, 2}^2 \frac{\delta t^2}{\delta x^2} -1 \right) + \delta t^2 V_{1, \;2} (r(x)) \right] \Psi_{1,\;2}(t, x)  \nb\\
&& + \mathfrak{v}_{1, 2}^2 \frac{\delta t^2}{\delta x^2} \left[ \Psi_{1,\;2}(t, x-\delta x) + \Psi_{1,\;2}(t, x+\delta x) \right] -\delta t^2  Q_{1, \;2} (r(x)) \Psi_{2,\;1}(t, x),
\eqn
\end{widetext}
where $\delta t$ and $\delta x$ are the step sizes for the $t$ and $x$ directions, respectively. They will be assigned suitable values in practice according to our usage. Here, $\mathfrak{v}_{1, 2}^2$ denote the speed factor in front of the $\partial^2/\partial x^2$ terms. Of course, for PDEs like Eqs.\eqref{master3a} and \eqref{master3b} we have $\mathfrak{v}_{1, 2}=1$. 

The calculations of $\Psi_{1,\;2}(t, x)$ will be performed on a isosceles triangular lattice in the $t-x$ Cartesian coordinate system. The bottom side of the  triangle contains $2N+1$ points so that $x/\delta x \in[-N, N ] \cap \mathbb{Z}$, where $N$ is a positive integer that will be chosen properly according to our usage.  The height of the triangle contains $N+1$ points so that $t/\delta t \in [1, N+1] \cap \mathbb{Z} $.
After $2N^2$ iterations, and with apt initial conditions, the functions $\Psi_{1,\;2}(t, x=0)$ could be obtained numerically. 
In practice, the initial conditions are chosen to be \cite{XinLi2020, Kai2016}
\bqn
\lb{FDM3}
 \Psi_{1,\;2}(t=0, x) &=& e^{-(x-1)^2/2}, \nb\\
 \left. \frac{d \Psi_{1,\;2}(t, x)}{dt} \right|_{t=0} &=& 0.
\eqn


\begin{figure*}[tbp]
\centering
	\begin{tabular}{c c c}
		\includegraphics[width=5.5cm]{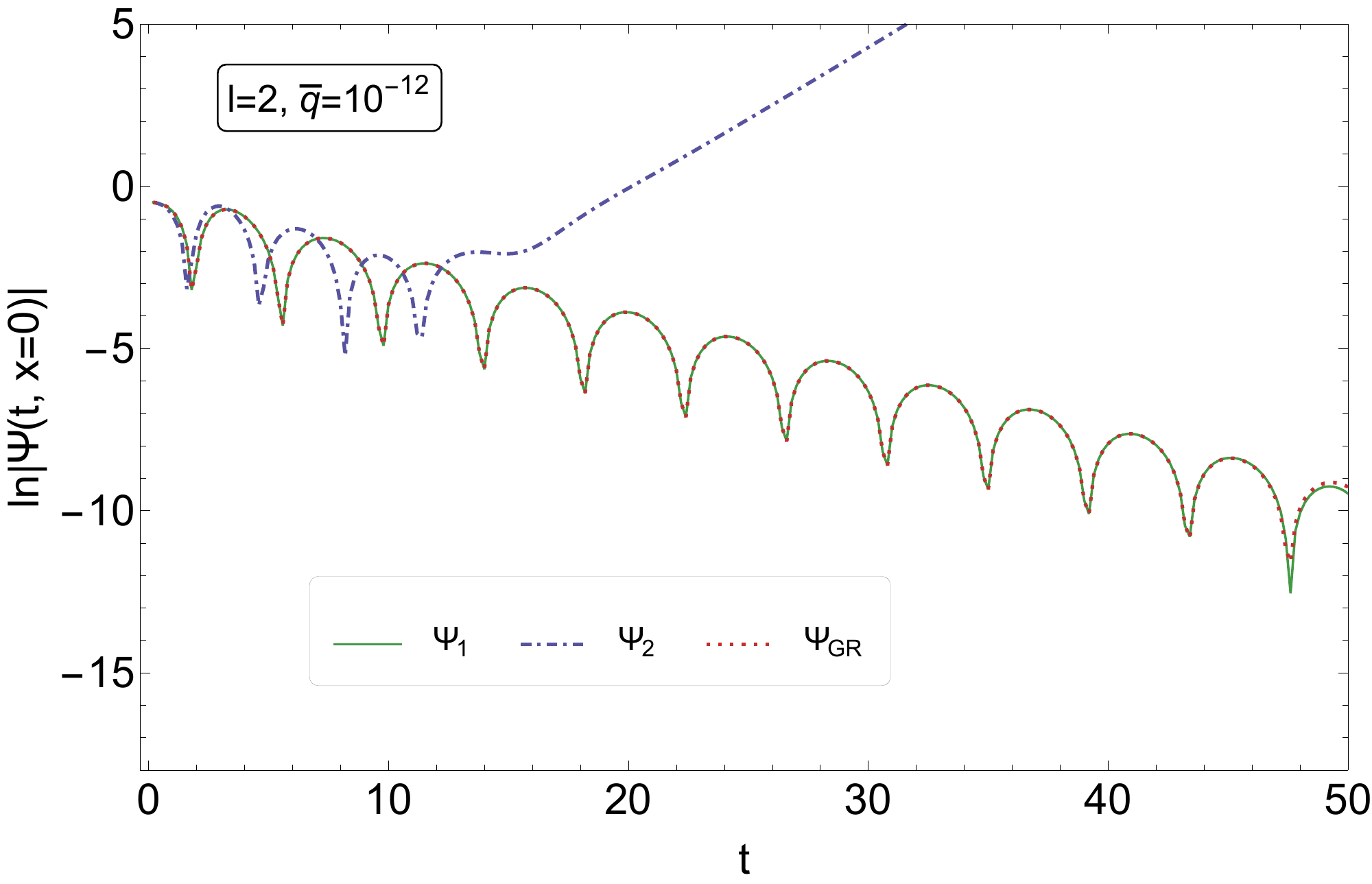} &   \includegraphics[width=5.5cm]{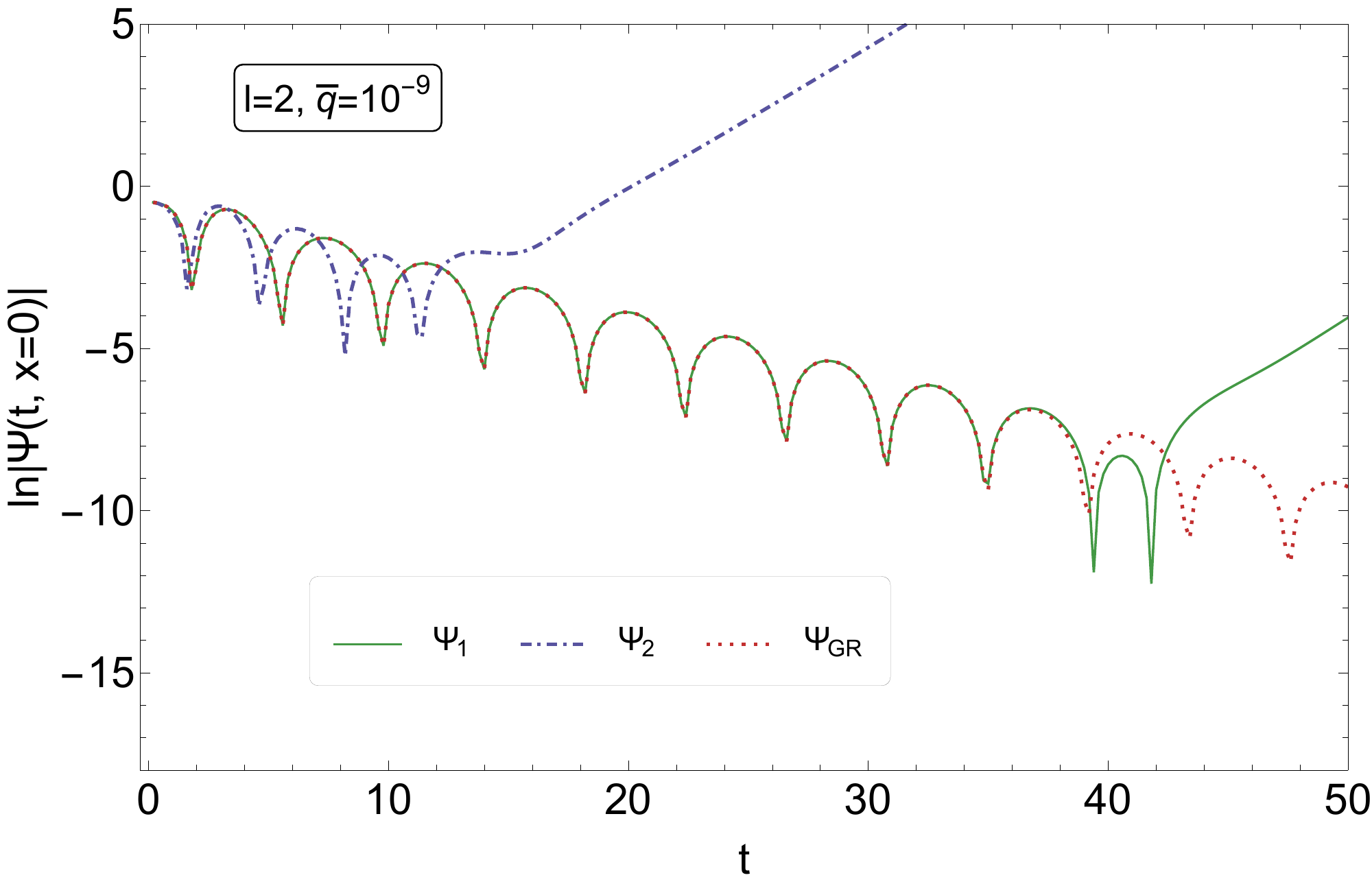} &   \includegraphics[width=5.5cm]{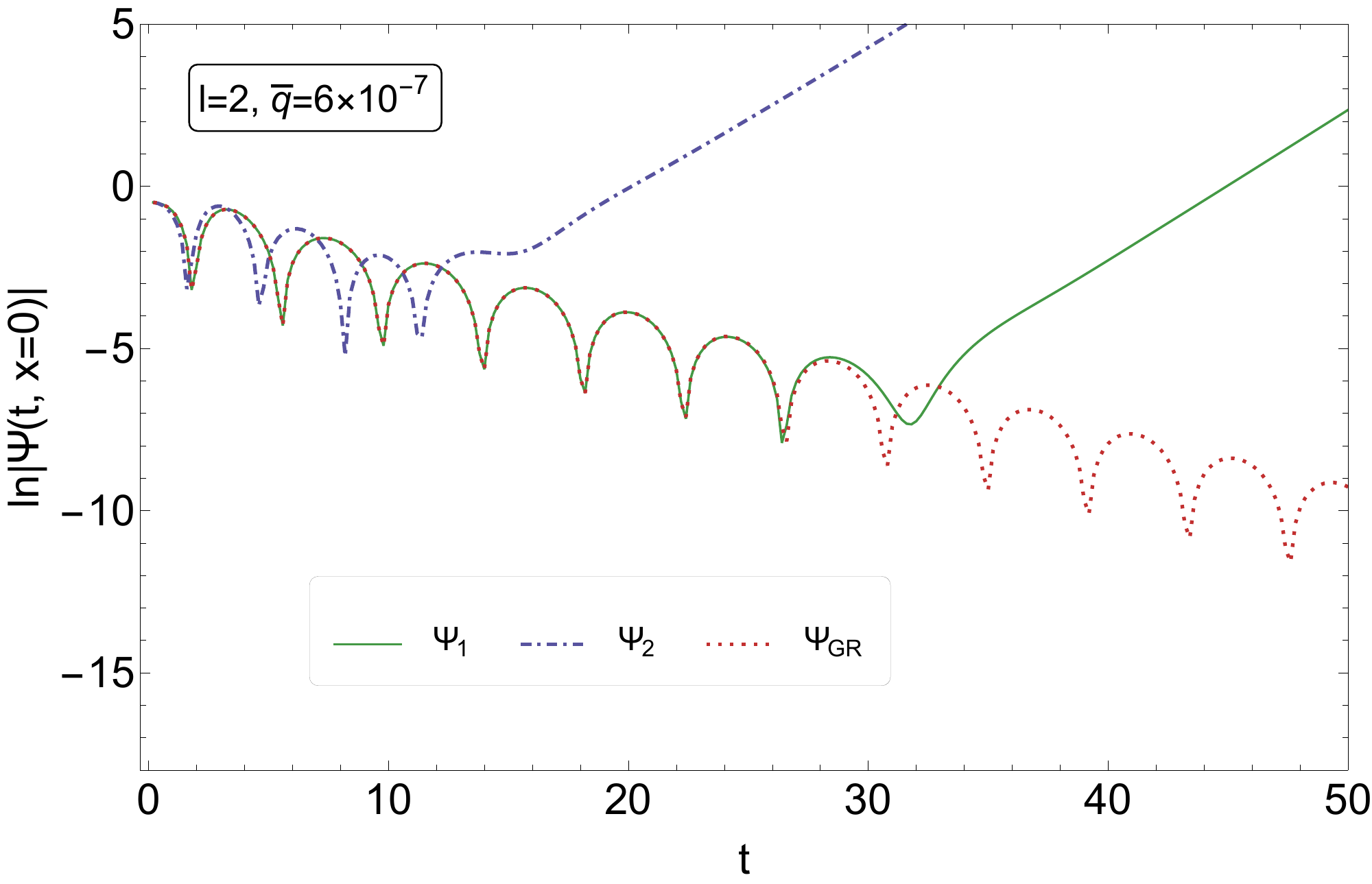} \\
		(a) & (b)  & (c)  \\[6pt]
  		\includegraphics[width=5.5cm]{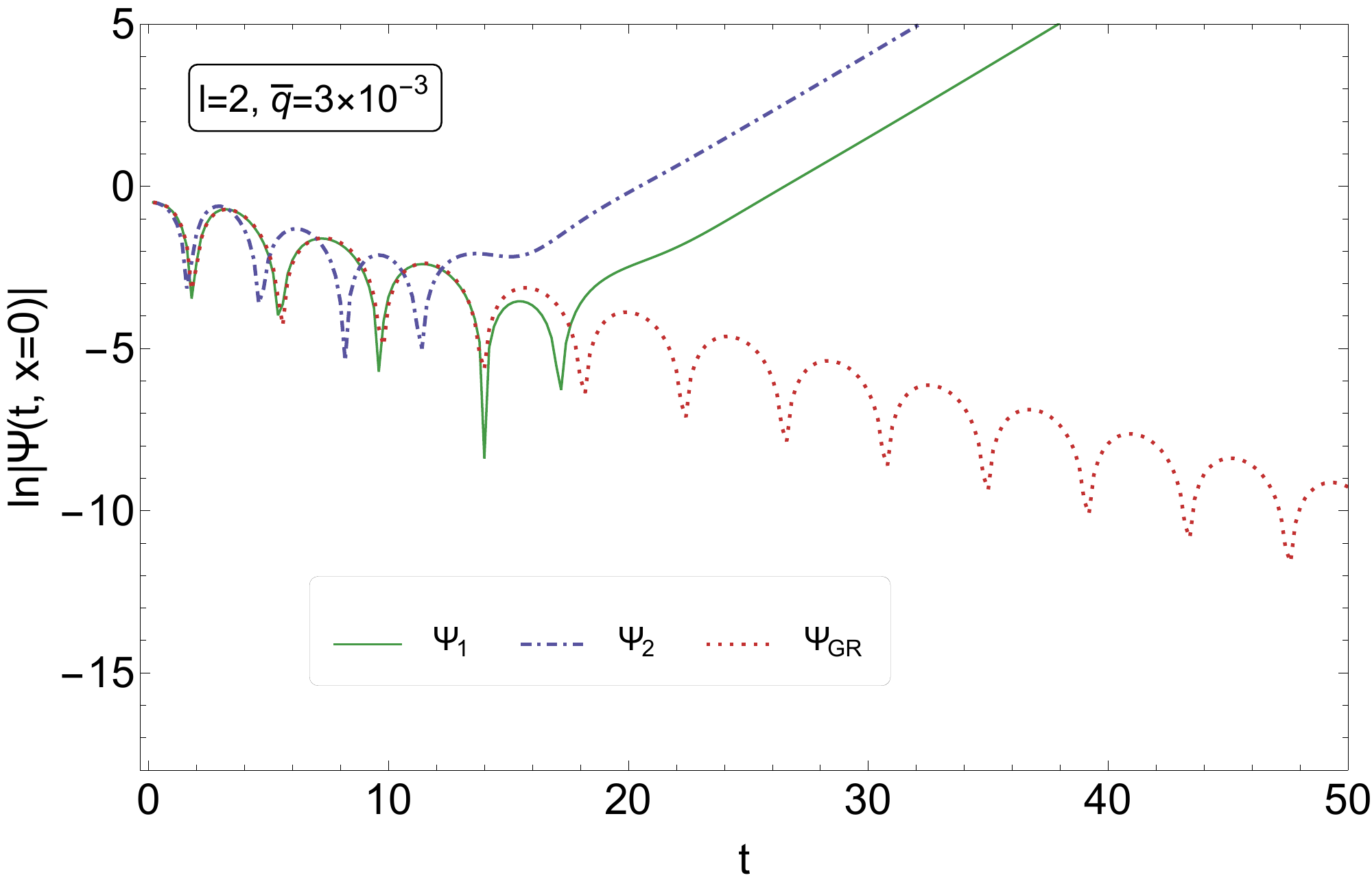} &   \includegraphics[width=5.5cm]{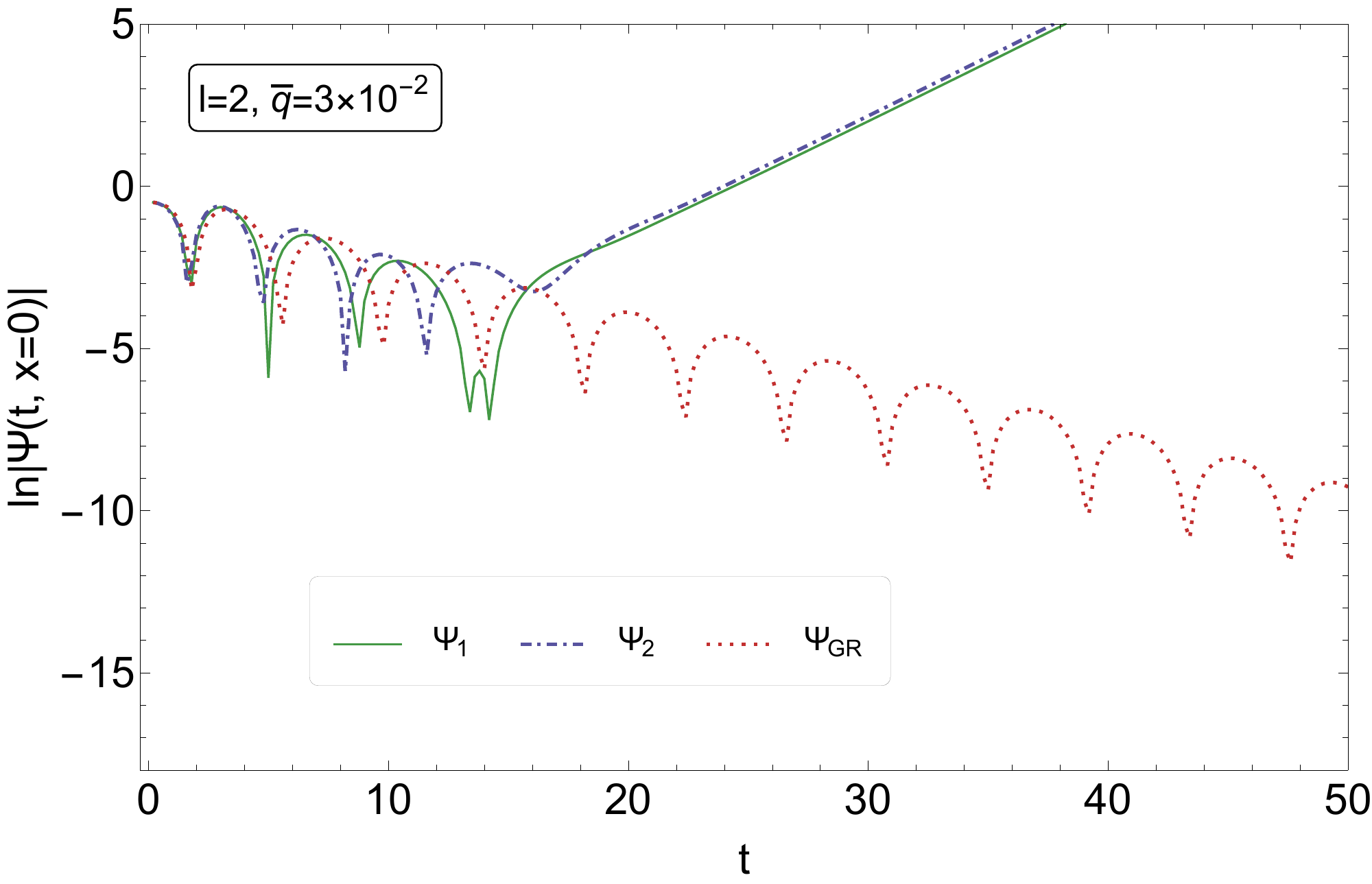} &   \includegraphics[width=5.5cm]{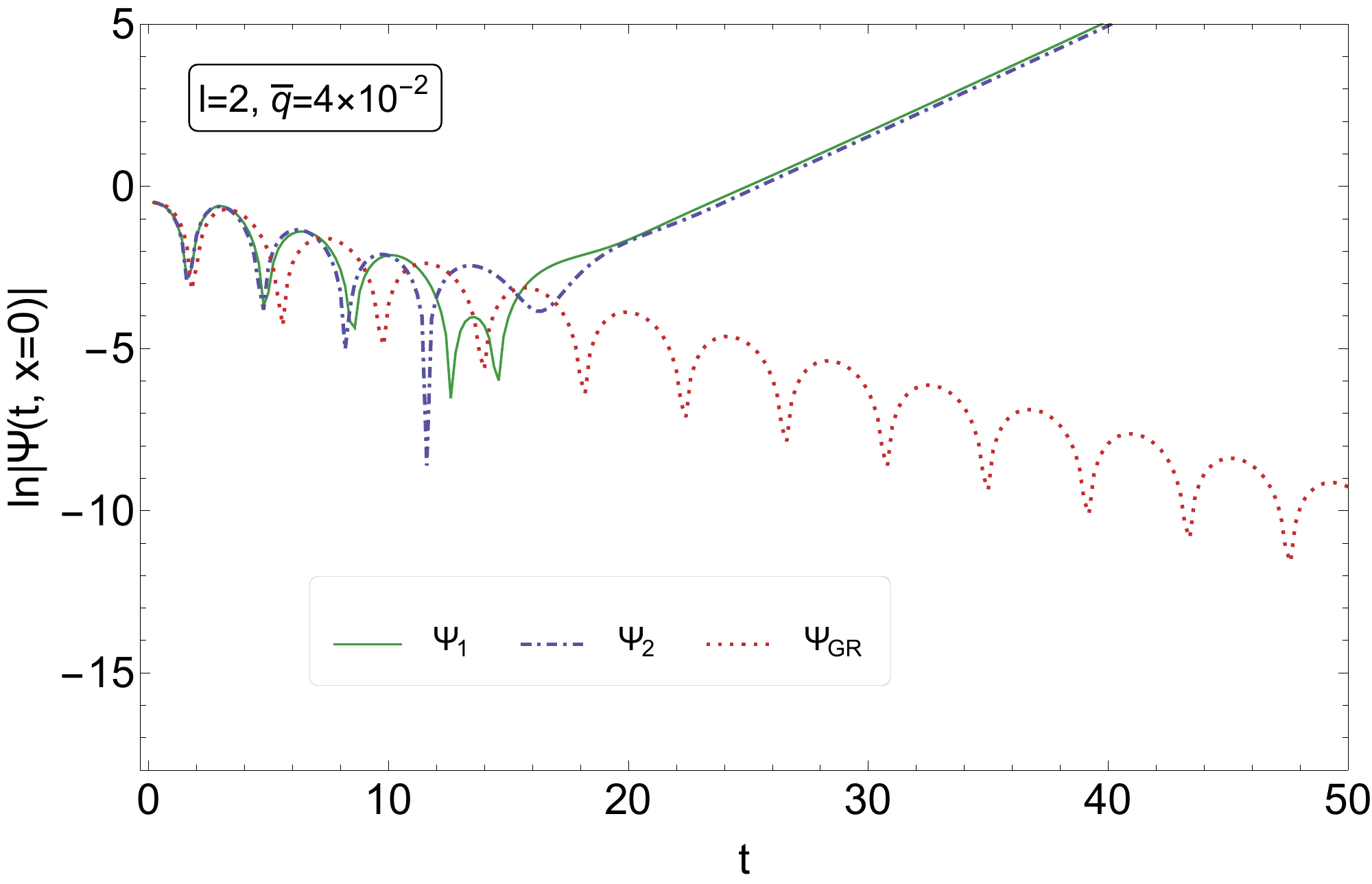} \\
		(d) & (e)  & (f)  \\[6pt]
  		\includegraphics[width=5.5cm]{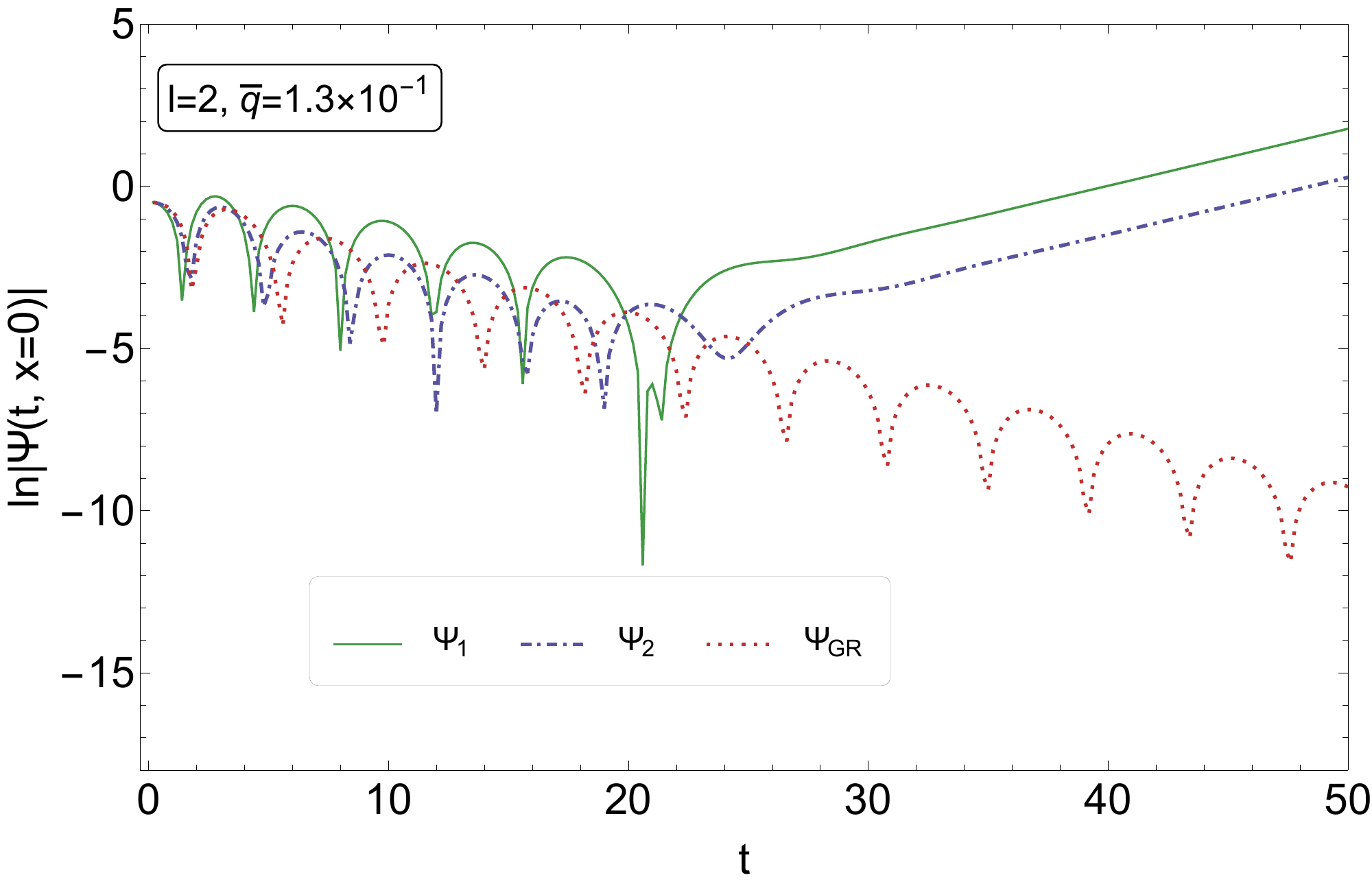} &   \includegraphics[width=5.5cm]{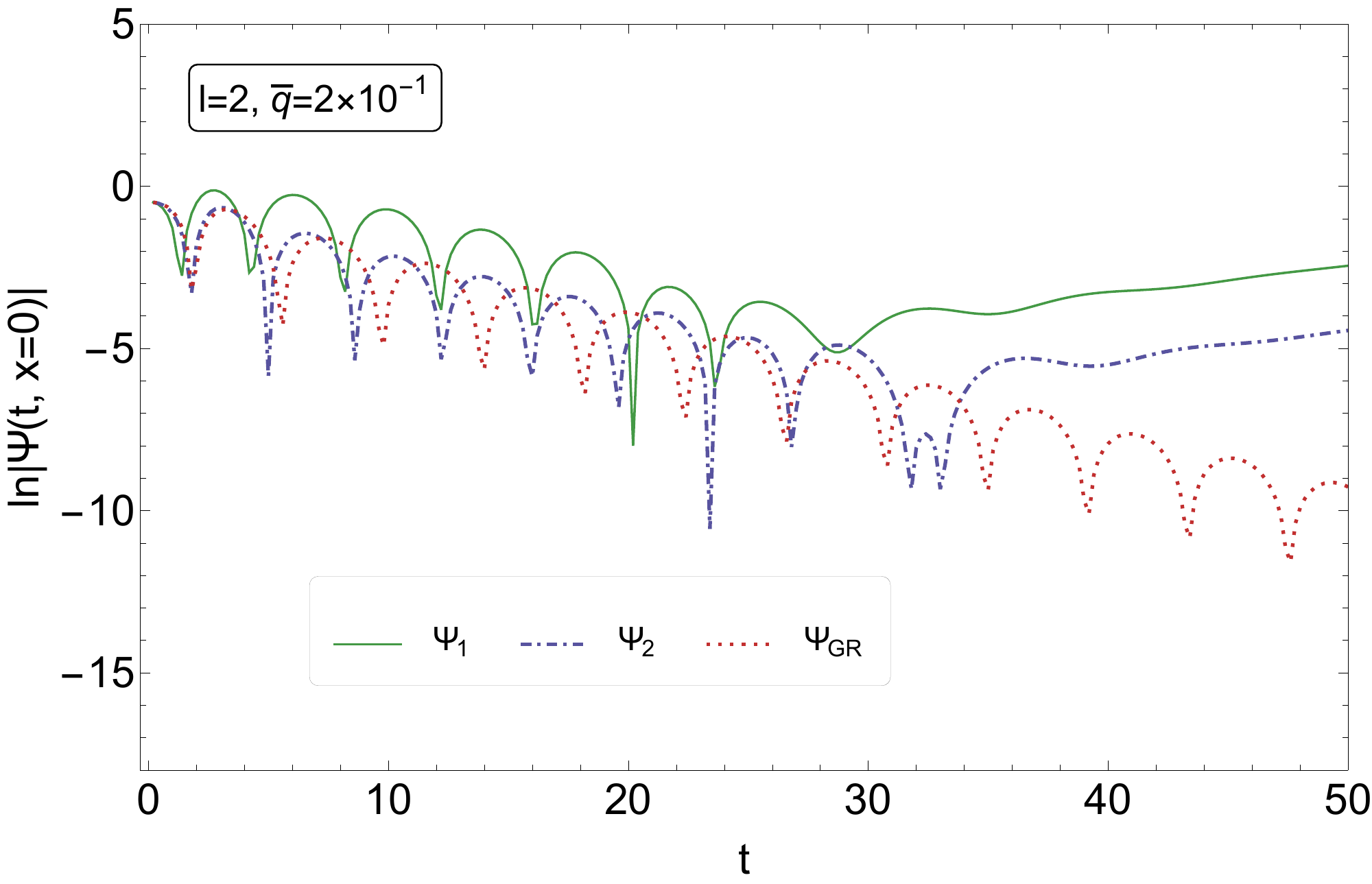} &   \includegraphics[width=5.5cm]{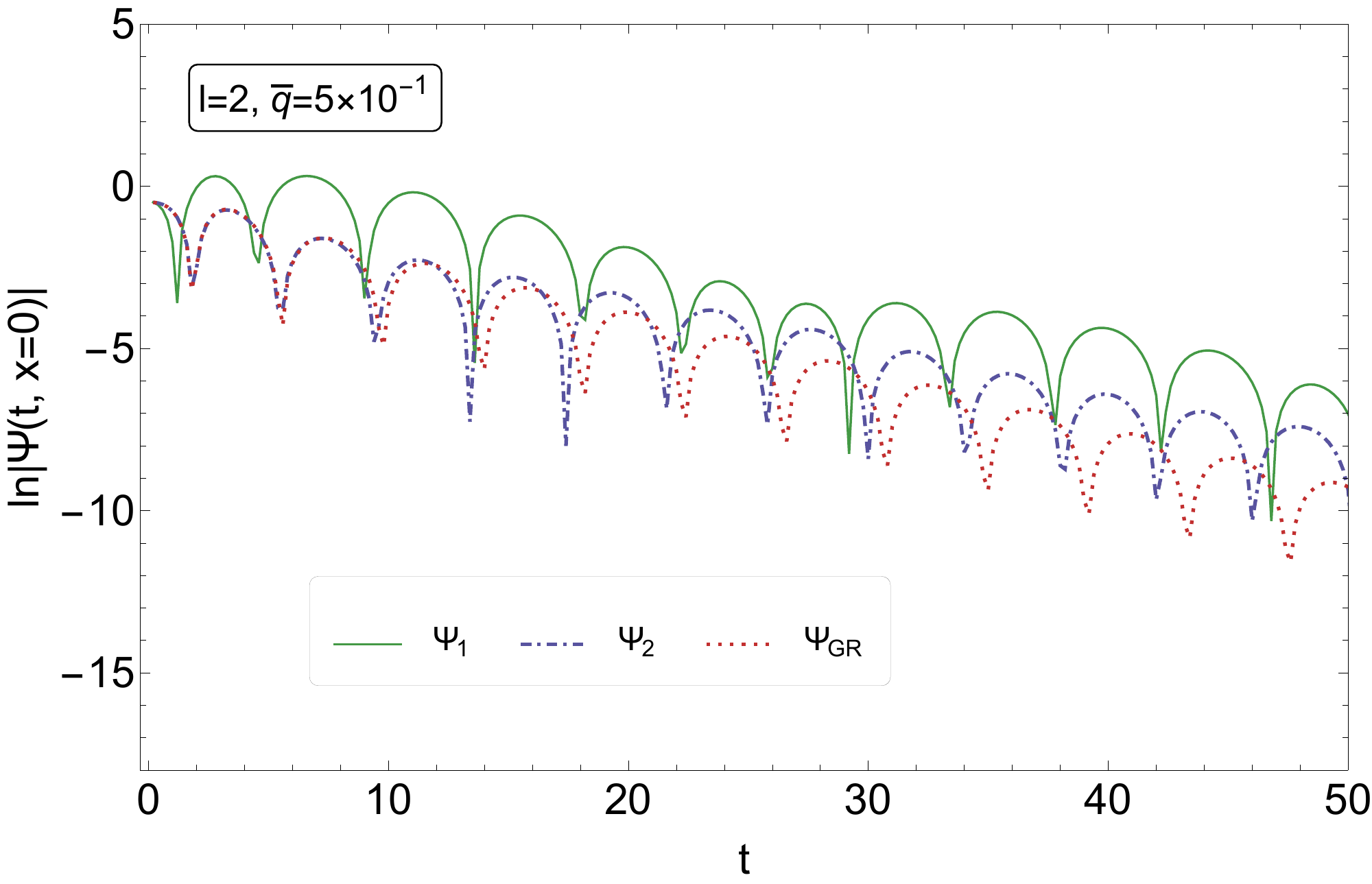} \\
		(g) & (h)  & (i)  \\[6pt]
	\end{tabular}
	\caption{The temporal evolution of the functions $\ln|\Psi_{1,2}(t, x=0)|$ for the $l=2$ mode and $c_{13}=c_4=0$ case, together with the GR case as a comparison ($\Psi_1$: green solid line; $\Psi_2$: purple dot-dashed line; $\Psi_{\text{GR}}$: red dotted line). Panels (a)-(i) correspond to ${\bar q}=10^{-12}, 10^{-9}, 6\times10^{-7}, 3\times10^{-3}, 3\times10^{-2}, 4\times10^{-2}, 3\times10^{-1}, 2 \times10^{-1}, 5\times10^{-1}$, respectively. Notice that, here we are adopting the unit system so that $c=G_N=2m=1$. Here $N$ is chosen to be 250.}
	\label{plot1}
\end{figure*}
\begin{figure}[htb]
	\includegraphics[width=\columnwidth]{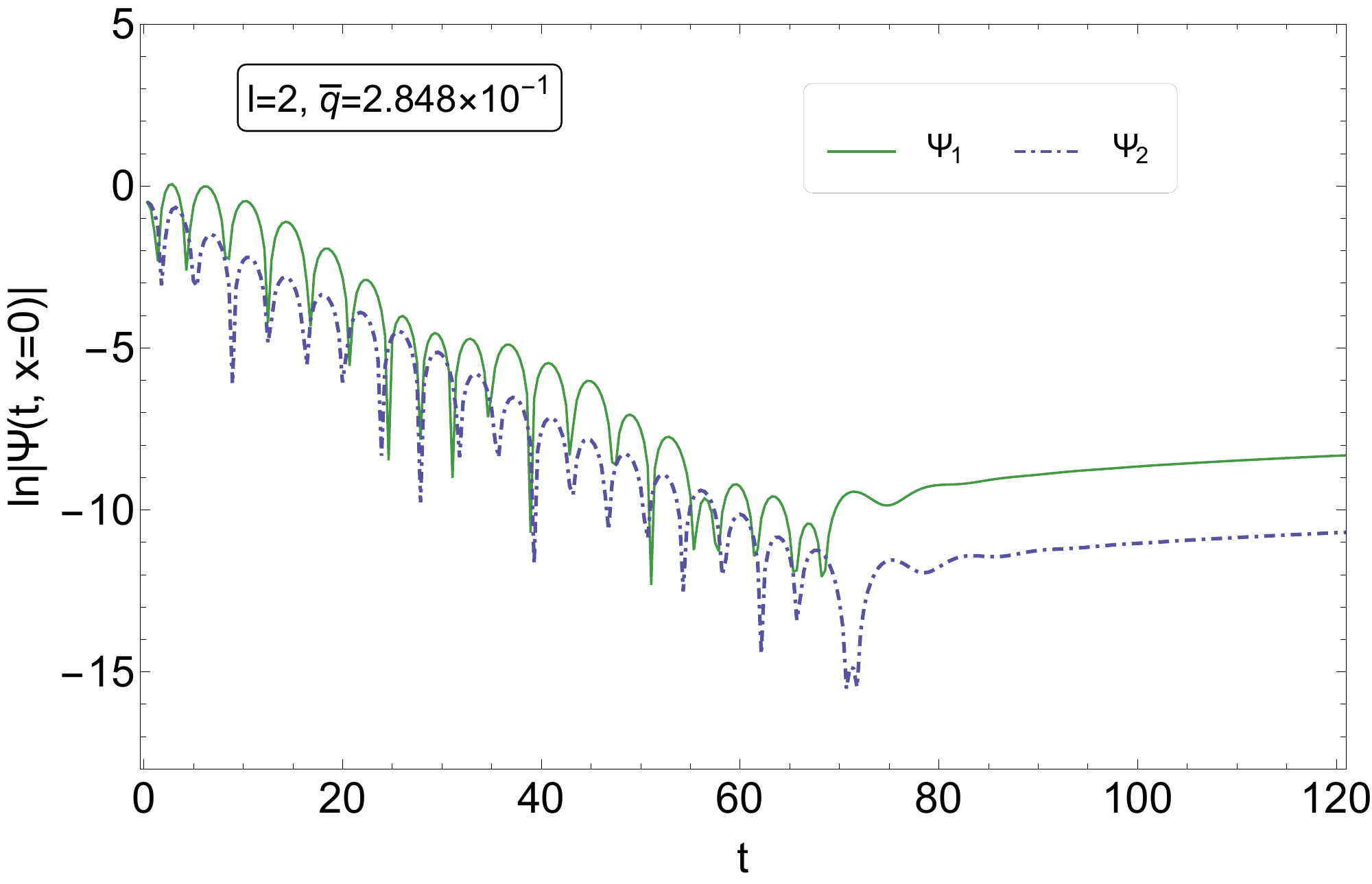} 
\caption{The temporal evolution of the functions $\ln|\Psi_{1,2}(t, x=0)|$ for the $l=2$ mode and ${\bar q}=2.848\times10^{-1}$, where we have set $c_{13}=c_4=0$. Notice that, here we are adopting the unit system so that $c=G_N=2m=1$. Here $N$ is chosen to be 350.} 
	\label{plot4}
\end{figure} 

Due to the observational significance of the $l=2$ mode \cite{Mag18, Shi2019, Ghosh2021, LVK2022}, we shall mainly focus on this case (Of course, the computations and analysis here could be easily extended to higher $l$'s). The main results of the corresponding solutions are exhibited in Fig. \ref{plot1}. In there we plot out the temporal evolution of  $\ln|\Psi_{1,2}(t, x=0)|$, together with the GR case as a comparison. Recall that we are adopting the unit system so that $c=G_N=2m=1$. To show it more explicit how the magnitude of ${\bar q}$ influences the behaviors of $\Psi_{1,2}$, we considered both physically allowed and forbidden ${\bar q}$'s. Based on Fig. \ref{plot1}, we have the following comments:

\begin{itemize}

\item  From panel (a) to panel (i), the value of ${\bar q}$ is increasing. We observed that, in general, the behavior of $\Psi_{1}$ is quite different from that of $\Psi_{2}$, which makes sense since the former is mainly from the gravitational perturbations [the contributions from the \ae{}ther field to $\Psi_1$ is suppressed by the small factor {$c_{1}$,} as seen from the definition \eqref{chi}] while the latter represents the contributions of the \ae{}ther field. 

\item  When $\bar q$ is small enough [e.g., panel (a)], the curve of $\Psi_{1}$ is almost overlapped with that of GR. In contrast, when $\bar q$ is large [e.g., panel (i)], the curve of $\Psi_{1}$ will deviate a lot from that of GR. Of course, these are what we expected.

\item As $\bar q$ approaching $0$ [cf., panels (a)-(d)], the deformation on the curve of  $\Psi_{2}$ tends to disappear. This makes sense since $V_2$ and $Q_2$ are of the form ${\cal O}(1)+{\cal O}({\bar q})$ for a tiny $\bar q$ [cf., \eqref{V1V2}].

\item In panels (d)-(i), physically forbidden $\bar q$'s were used [cf., Eq.\eqref{c1234a}]. Even though, from there we see clearly how the dynamical instability arises. When $\bar q$ is large enough,  $\Psi_{1,2}$ behaves in a healthy way [e.g., panel (i)]. However, for small enough $\bar q$'s,  $\Psi_{1,2}$ will finally blow up, which reflects the existence of a dynamical instability \cite{Konoplya2018} [e.g., panel (g)]. As could be estimated, the critical point should occur at 
{$\bar q \simeq 0.3$} [cf., panels (h) and (i)].

Since the position of the critical point has special significance, we carry out a more careful investigation of that. By selecting a $\bar q$ near the critical point, the curves of $\Psi_{1, 2}$ are shown in Fig. \ref{plot4} with a larger scope of $r$. In there we observe the plateaux appearing on the curves of $\Psi_{1,2}$, which, from the phenomenological point of view, indicates that the current $\bar q$ is around the critical point \cite{Kai2016}.

\item In contrast, physically allowed $\bar q$'s [cf., Eq.\eqref{c1234a}] were selected for panels (a)-(c). Although  $\Psi_{1,2}$ will blow up soon or later, we observe that the curves of  $\Psi_{1}$'s are almost overlapped with that of GR, before blowing up. That makes sense since those physically allowed $\bar q$'s are extremely small, and ${\bar q}=0$ means the GR limit. 

\item From panels (a)-(h), we noticed that the times of blowing up for $\Psi_{1,2}$ are getting earlier and earlier simultaneously as $\bar q$ getting smaller from a big enough value [cf., panel (h)]. This tendency vaporizes at ${\bar q} \approx 3\times 10^{-2}$, where $\Psi_{1}$ and $\Psi_{2}$ exchange their chronological order of blowing up [cf., panels (e) and (f)]. Starting from this point, the curve of $\Psi_{2}$ is getting frozen step by step [cf., panels (d) and (e)], and tends to lose its sensitivity on the magnitude of $\bar q$ [cf., panels (a)-(d), as mentioned earlier]. While for the curve of $\Psi_{1}$, the time of its blowing up is getting  more and more postponed as  $\bar q$ approaches zero [cf., panels (a)-(e)]. 

\end{itemize}

\begin{figure}[htb]
	\includegraphics[width=\columnwidth]{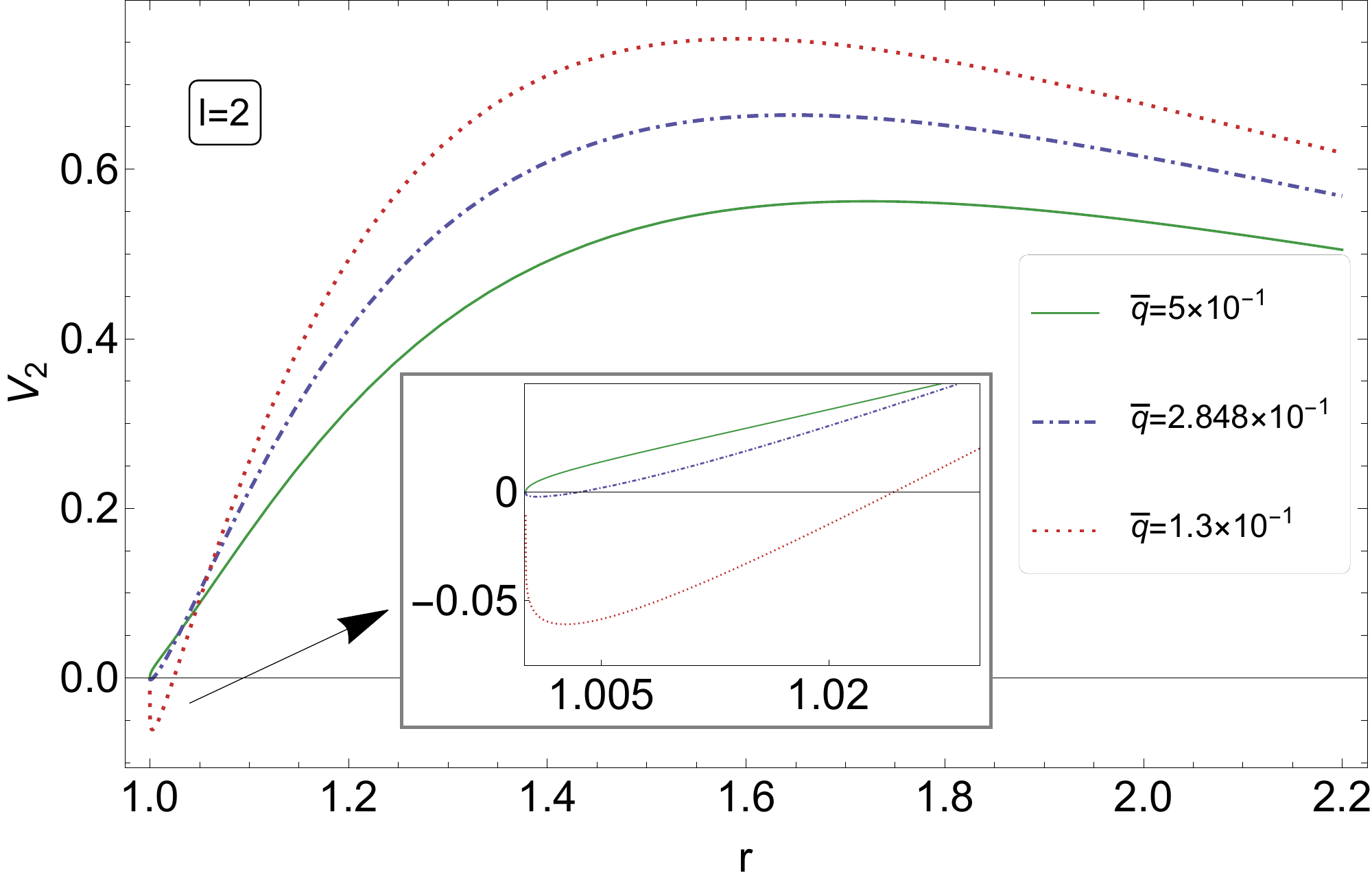} 
\caption{The effective potential $V_2$ [cf., Eq.\eqref{V1V2}] as a function of $r$ for the $l=2$ mode and different choices of $\bar q$'s, where we have set $c_{13}=c_4=0$. Notice that, here we are adopting the unit system so that $c=G_N=2m=1$.} 
	\label{plot3}
\end{figure} 

Qualitatively speaking, the dynamical instability \footnote{One should avoid confusing this kind of instability with the ghost and Laplacian instabilities discussed in \cite{Tsujikawa21}. In fact, by following the formalism of \cite{Tsujikawa21} and using the Lagrangian represented by \eqref{Lodd}, it's straightforward to exclude these two kinds of instabilities for the $c_{13}=c_4=0$ and ${\bar q} > 0$ cases.} mentioned above could mainly be attributed to the features of the effective potential $V_2$. As seen from Fig. \ref{plot3}, the effective potential $V_2$ can stay positive once $\bar q$ is large enough. In contrast, a small $\bar q$ may drag a part of $V_2$ to the negative regime. 
By following the formalism of \cite{Takahashi2013, Gannouji2022}, we notice that being negative on $V_2$ (or $V_1$, although we abbreviated this part of the discussion since $V_1$ was observed to stay positive for all the reasonable $\bar q$'s, e.g., those considered in Fig. \ref{plot1}) can introduce instability to our dynamical system. This explains the peculiar behaviors of $\Psi_{1,2}$ shown in Fig. \ref{plot1} and Fig. \ref{plot4} \footnote{Notice that, although $V_1(r)$ stays positive (at least for the physically allowed values of  $\bar q$'s), a general  analysis to the PDEs tells us that the blowing-up feature of $\Psi_2$ could be conveyed to $\Psi_1$ \cite{Evans2010}, as now  $\Psi_2$ is acting as its source term [cf., Eq.\eqref{master3a}]. On the other hand, one should not forget that $\Psi_1$ in fact carries the information of $\Psi_2$ by definitions \eqref{chi} and \eqref{q1q2}.}. Thus, although it sounds a little counter-intuitive, we conclude that a relatively large $\bar q$ can guarantee the stability of our dynamical system \footnote{Similar analysis to this kind of dynamical instability could be done for \cite{Kai2016}. It is also due to the negative sign of the corresponding effective potential (under certain choices of the coupling constant), that instability occurs. Nonetheless, different from our case, in there, a smaller coupling constant could erase such an instability.},  while on the other hand (as expected) enhancing the deviation between $\Psi_1$ and $\Psi_{\text{GR}}$, and vice versa.

\begin{figure}[htb]
	\includegraphics[width=\columnwidth]{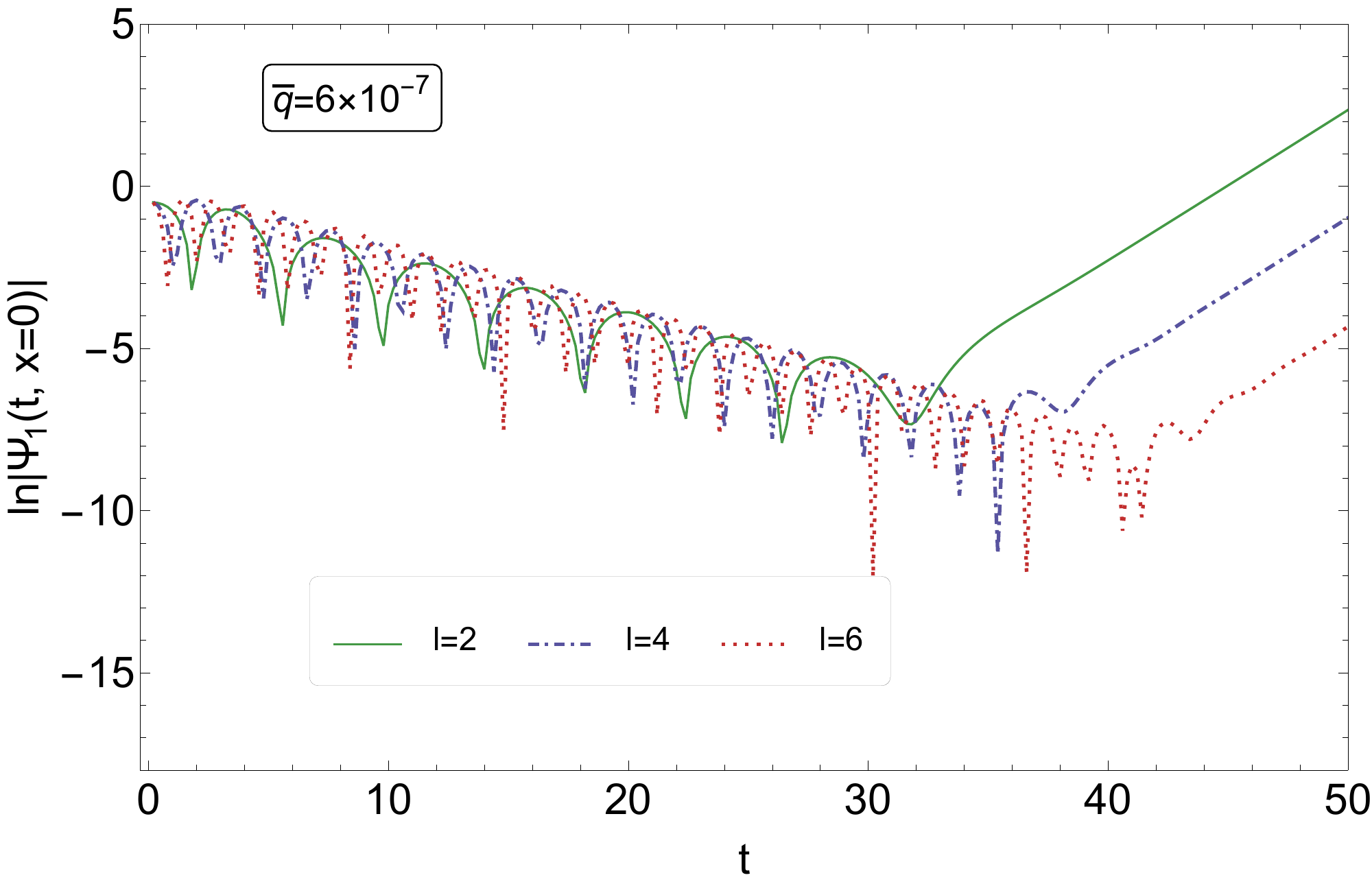} 
 	\includegraphics[width=\columnwidth]{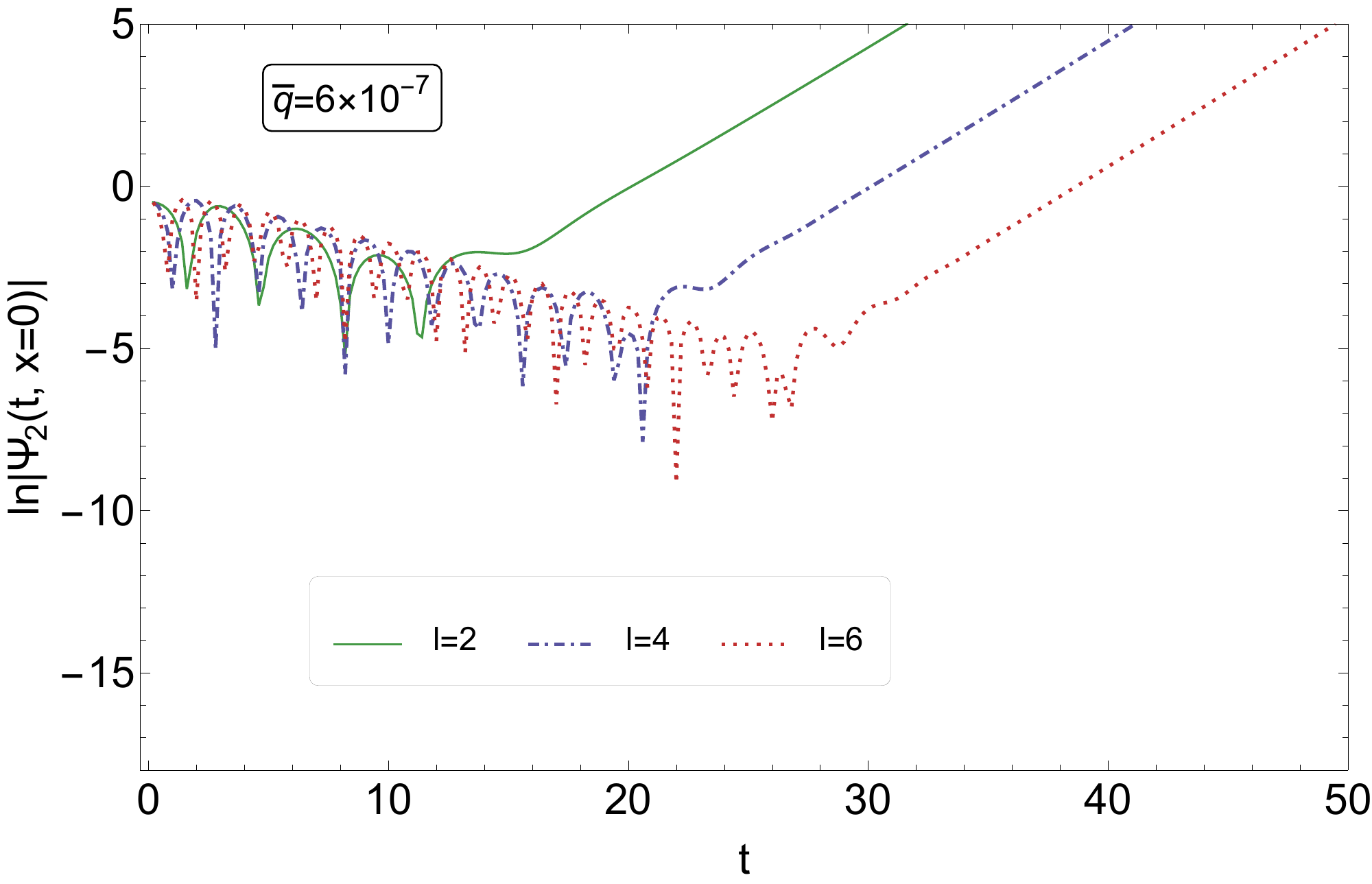} 
\caption{The temporal evolution of the functions $\ln|\Psi_{1}(t, x=0)|$ (upper panel) and $\ln|\Psi_{2}(t, x=0)|$ (lower panel)  for the $l=2, 4, 6$ modes (correspond to the green solid line, purple dot-dashed line and red dotted line, respectively) and ${\bar q}=6 \times10^{-7}$, where we have set $c_{13}=c_4=0$. Notice that, here we are adopting the unit system so that $c=G_N=2m=1$. Here $N$ is chosen to be 250.} 
	\label{plot6}
\end{figure}

{Another interesting phenomenon is that, the instability mentioned above could be relaxed a little bit for a larger $l$. As seen from Fig.\ref{plot6}, for same $\bar {q}$, choosing a relatively larger $l$ could postpone the time of being blow-up for $\Psi_{1,2}$ (what we observed here is different from that of \cite{Konoplya2008}). That is, higher multipoles tend to be more stable.
 }

 It is worth mentioning here that, the ratio between $\delta x$ and $\delta t$ plays an important role in controlling the numerical stability during the calculations of  $\Psi_{1,2}$. By defining $\mathfrak{k} \equiv \delta x/\delta t$, a Courant-like stability condition is given by (see, e.g., the \S 6.5 of \cite{Habermanb} and \cite{LeVeque2007})
\bqn
\lb{stable1}
\frac{2 \mathfrak{v}_{1, 2}^2}{\mathfrak{k}^2} {\bar k}- \delta t^2 \left( V_{1, \;2} + Q_{1, \;2} \right) \in (-4, 0),
\eqn
where ${\bar k} \in (-2, 0)$. Since the numerical values of $ V_{1, \;2}$ and $ Q_{1, \;2}$ can be easily obtained, we can solve for the approximate valid region of $\mathfrak{k}$ from \eqref{stable1} for any specific $\delta t$. We then noticed that, for the $l=2$ case and a typical $\delta t$ [viz., $\delta t \in (0, 1)$], $\mathfrak{k}=1.2$ could guarantee the condition \eqref{stable1}. This could be further confirmed by monitoring the behavior of solved $\Psi_{1,2}$ with the varying $\mathfrak{k}$. Indeed,  we observed that the deformation of curves of $\Psi_{1,2}$ shows a convergent behavior for $ \mathfrak{k} \gtrsim 1$ [In contrast, the curve of resultant  $\Psi_{1}\setminus\Psi_{2}$ is sensitive to the choice of  $\mathfrak{k}$ for  $\mathfrak{k} \in (0, 1)$, which reflects a numerical instability]. Notice that, such a convergent behavior also supports the validity of the individual values of our chosen $\delta t$ and $\delta x$ \footnote{ { To be specific, the emergence of such a convergent behavior means the reasonable variations of  $\delta t$ and $\delta x$ won't change our numerical results. Technically speaking,  $\delta t$ and $\delta x$ are the most significant factors in our numerical method. Other software-related factors were found to have trivial influence on the numerical results, as long as we are working under a hing enough precision and accuracy (this is exactly what we did in practice). Therefore, by following some general protocols of judging a numerical method, we are able to exclude the numerical instabilities for our results. } }.


\section{Conclusions}
 \renewcommand{\theequation}{5.\arabic{equation}} \setcounter{equation}{0}

In this paper, we investigated QNMs of odd-parity perturbations in Einstein-\AE{}ther theory. Specially, here we pay attention to a kind of wormhole-like backgrounds, described by \eqref{backeqn2}. To find the corresponding QNMs, we first work in the isotropic coordinate and simplify the 2nd-order Lagrangian to Eq.\eqref{Lodd} by following \cite{Tsujikawa21}. Then, the desired master equations in the Schwarzschild coordinate could be obtained in terms of a set of coupled PDEs [cf., Eqs.\eqref{master3a} and \eqref{master3b}], {which depend only on the coupling constant $c_1$ (or equivalently, $\bar q$), after setting $c_{13} = c_4 = 0$ from the requirement of no-ghost conditions \cite{Tsujikawa21} and constraints from observations [cf., \eqref{c1234ac}].} 

As has been mentioned in Sec. I, there are many different techniques in general for calculating QNMs. However, in reality, the FDM is identified as one of the most apt ways for our case. By mainly focusing on the $l=2$ mode (due to its physical significance \cite{Mag18, Shi2019, Ghosh2021, LVK2022}), that set of PDEs are solved numerically with the help of the recursion formula \eqref{FDM2}. By varying the coupling constant {$c_{1}$,} or equivalently, $\bar q$ [cf., \eqref{backeqn2}], the corresponding solutions are shown in terms of $\Psi_{1, 2}$ vs. $t$, and are exhibited in Figs. \ref{plot1} and \ref{plot4}. 

As shown in Sec. IV. B, we read off the features of  $\Psi_{1, 2}$, and their dependence on $\bar q$ from  Figs. \ref{plot1} and \ref{plot4}. Specially, we found that choosing the physically allowed values of  $\bar q$'s [cf.,  \eqref{c1234ac}] will bring the dynamical system to a kind of instability (which is different from that found in \cite{Tsujikawa21}), as observed from the panels (a)-(c) of Fig. \ref{plot1}. This phenomenon is actually consistent with the behaviors of the effective potentials [cf., Eq.\eqref{V1V2}], since a physically allowed value of $\bar q$ will drag a part of the potential $V_2$ to the negative regime [cf., Fig. \ref{plot3}]. Therefore, we conclude that the background solutions described by   \eqref{backeqn2} are not stable against the odd-parity perturbations and should be ruled out. 



\section*{Acknowledgements}
This work is supported by the National Key Research and Development Program of China under Grant No.2020YFC2201503, the National Natural Science Foundation of China under Grant No. 12275238, No. 11975203, No. 11675143, No. 12205254, the Zhejiang Provincial Natural Science Foundation of China under Grant No. LR21A050001 and LY20A050002, and the Fundamental Research Funds for the Provincial Universities of Zhejiang in China under Grant No. RF-A2019015.




\section*{Appendix A: Integral formulas for spherical harmonics}
\renewcommand{\theequation}{A. \arabic{equation}} \setcounter{equation}{0}

The spherical harmonics $Y_{lm} (\theta, \phi)$ has the following integral features \cite{Zettili, Kase:2018voo}
\begin{widetext}
 \bqn
\lb{Ylm1}
&& \int_0^{2 \pi} d \phi \int_0^{\pi} d \theta Y_{l0}^2(\theta, \phi) \sin \theta = 1, \\ \nb\\
&& \int_0^{2 \pi} d \phi \int_0^{\pi} d \theta \left| \frac{\partial}{\partial \theta} Y_{l0}(\theta, \phi) \right|^2 \sin \theta = L, \\ \nb\\
&& \int_0^{2 \pi} d \phi \int_0^{\pi} d \theta \Bigg[ \csc \theta \left| \frac{\partial}{\partial \theta} Y_{l0}(\theta, \phi) \right|^2  
+ \sin \theta \left| \frac{\partial^2}{\partial \theta^2} Y_{l0}(\theta, \phi) \right|^2 \Bigg] = L^2. ~~~
\eqn
\end{widetext}


\section*{Appendix B: Expressions of $\beta_i$}
\renewcommand{\theequation}{B. \arabic{equation}} \setcounter{equation}{0}
\begin{widetext}
The coefficients $\beta_n$ that appear in Eq.(\ref{Lodd}) are given by
\bqn
\lb{beta1}
\beta_1 &\equiv&  -\frac{L e^{-2 (2 \mu +\nu )}}{2 \left[c_{14} \rho ^2 \left(\mu '\right)^2+4 \rho  \mu ' \left(\rho  \nu '+1\right)-L+2 \rho ^2 \left(\nu '\right)^2+4 \rho  \nu '+2\right]}, \nb\\
\beta_2  &\equiv&   \frac{c_{14} L e^{-2 (\mu +\nu )}}{\rho ^2},\quad
\beta_3  \equiv   \frac{L e^{-2 (\mu +2 \nu )}}{4-2 L},\quad
\beta_4  \equiv   -\frac{c_1 L e^{-4 \nu }}{\rho ^2},\nb\\
\beta_5  &\equiv&  -\frac{L e^{-2 (\mu +2 \nu )} \left[\left(c_{14}+4\right) \left(-\rho ^2\right) \left(\mu '\right)^2-4 \rho  \mu ' \left(3 \rho  \nu '+2\right)+4 L+10 \rho ^2 \left(\nu '\right)^2+24 \rho  \nu '\right]}{8 (L-2) \rho ^2},\nb\\
\beta_6  &\equiv&   -\frac{L e^{-4 \nu } \left[-2 \left(c_1-c_{14}\right) \rho  \mu ' \left(\rho  \nu '+1\right)+\left(c_1+2 \left(c_{14}-1\right) c_{14}\right) \rho ^2 \left(\mu '\right)^2+c_1 L\right]}{\rho ^4},\nb\\
\beta_7  &\equiv&   \frac{2 c_{14} L e^{-\mu -4 \nu } \mu '}{\rho ^2}.
\eqn
Here, a prime in the superscript denotes the derivative with respect to $\rho$.
\end{widetext}


\section*{Appendix C: Expressions of $\gamma_{ij}$}
\renewcommand{\theequation}{C. \arabic{equation}} \setcounter{equation}{0}

\begin{widetext}
The coefficients $\gamma_{ij}$ that appear in Eq.(\ref{master1}) are given by
\bqn
\lb{gamma1}
\gamma_{11} &\equiv&  -\frac{e^{2 \mu -2 \nu } \left(c_{14} \rho ^2 \left(\mu '\right)^2+4 \rho  \mu ' \left(\rho  \nu '+1\right)-L+2 \rho ^2 \left(\nu '\right)^2+4 \rho  \nu '+2\right)}{L-2},\nb\\
\gamma_{12} &\equiv&  \frac{e^{2 \mu -2 \nu } \left(\rho  \mu '+\rho  \nu '-2\right) \left(c_{14} \rho ^2 \left(\mu '\right)^2+4 \rho  \mu ' \left(\rho  \nu '+1\right)-L+2 \rho ^2 \left(\nu '\right)^2+4 \rho  \nu '+2\right)}{(L-2) \rho },\nb\\
\gamma_{13} &\equiv&  \frac{e^{2 \mu -2 \nu } \left(c_{14} \rho ^2 \left(\mu '\right)^2+4 \rho  \mu ' \left(\rho  \nu '+1\right)-L+2 \rho ^2 \left(\nu '\right)^2+4 \rho  \nu '+2\right)}{4 (L-2) \rho ^2} \nb\\
&& \times  \left(\left(c_{14}+4\right) \left(-\rho ^2\right) \left(\mu '\right)^2-4 \rho  \mu ' \left(3 \rho  \nu '+2\right)+4 L+10 \rho ^2 \left(\nu '\right)^2+24 \rho  \nu '\right),\nb\\
\gamma_{14} &\equiv&  -\frac{2 c_{14} e^{3 \mu -2 \nu } \mu ' \left(c_{14} \rho ^2 \left(\mu '\right)^2+4 \rho  \mu ' \left(\rho  \nu '+1\right)-L+2 \rho ^2 \left(\nu '\right)^2+4 \rho  \nu '+2\right)}{\rho ^2},\nb\\
\gamma_{21}  &\equiv&  \frac{c_1 e^{2 \mu -2 \nu }}{c_{14}},\quad
\gamma_{22}  \equiv  \frac{c_1 e^{2 \mu -2 \nu } \left(\mu '-\nu '\right)}{c_{14}},\nb\\
\gamma_{23}  &\equiv&  -\frac{e^{2 \mu -2 \nu } }{c_{14} \rho ^2}  \Big[-2 \left(c_1-c_{14}\right) \rho  \mu ' \left(\rho  \nu '+1\right) \nb\\
&& +\left(c_1+2 \left(c_{14}-1\right) c_{14}\right) \rho ^2 \left(\mu '\right)^2+c_1 L\Big],\nb\\
\gamma_{24}  &\equiv&  e^{\mu -2 \nu } \mu '.
\eqn
Here, a prime in the superscript denotes the derivative with respect to $\rho$.
\end{widetext}

\section*{Appendix D: The formulas for obtaining Schr{\"o}dinger-like PDEs}
\renewcommand{\theequation}{D. \arabic{equation}} \setcounter{equation}{0}

The procedures for obtaining  Eqs.\eqref{master3a} and \eqref{master3b} may not be that straightforward. To show it more clearly, here we introduce the formulas used in this part of calculations.
Suppose we have a set of linearly coupled 2nd-order PDEs of the form
\bqn
\lb{Beqn1}
\left[{\eta}_1(r) \frac{\partial^2}{\partial r^2} +{\bar\eta}_1(r) \frac{\partial}{\partial r} +{\tilde \eta}_1(r)  - \frac{\partial^2}{\partial t^2}\right] J_1 = \xi_1(r) J_2(t,r),\nb\\
\left[{\eta}_2(r) \frac{\partial^2}{\partial r^2} +{\bar\eta}_2(r) \frac{\partial}{\partial r} +{\tilde \eta}_2(r)  - \frac{\partial^2}{\partial t^2}\right] J_2 = \xi_2(r) J_1(t,r).\nb\\
\eqn
Let us introduce a set of new functions $\{\Psi_1, \Psi_2\}$ and a new variable $x$ by
\bqn
\lb{Beqn2}
J_{1,2} = q_{1,2}(r) \Psi_{1,2}(t,x(r)), \quad \frac{dr}{dx} = p(r).
\eqn
Then, choosing  $q_{1,2}$ as
\bqn
\lb{Beqn3}
q_{1,2}(r) &=& \sqrt{p(r)} \exp{\left( -\frac{1}{2} \int \frac{ {\bar \eta}_{1,2}}{ {\eta}_{1,2}} dr \right)}, 
\eqn
we find that  Eq.\eqref{Beqn1} can be written as a set of coupled Schr{\"o}dinger-like PDEs
\bqn
\lb{Beqn4}
\left[\frac{\eta_{1,2}}{p^2} \frac{\partial^2}{\partial x^2} - \left( \frac{\partial^2}{\partial t^2} + V_{1,2}  \right) \right] \Psi_{1,2} = Q_{1,2}  \Psi_{2,1},
\eqn
where
\bqn
\lb{Beqn5}
V_{1,2}(r) &\equiv& -\Bigg\{ \frac{\eta_{1,2}}{4 p^2} \bigg[ 2 p \frac{d\left( p'-p {\bar \eta}_{1,2}/\eta_{1,2} \right)}{dr}  \nb\\
&& ~~~~ - \left( p'-p {\bar \eta}_{1,2}/\eta_{1,2} \right)^2 \bigg] + {\tilde \eta}_{1,2} \Bigg\}, ~~~~~~~~
\eqn
and
\bqn
\lb{Beqn6}
Q_{1,2}(r) &\equiv& \xi_{1,2} \frac{q_{2,1}}{q_{1,2}}.
\eqn
Here a prime denotes the derivative with respect to $r$. 
In addition, a common choice of $p(r)$ is   $p(r)=\sqrt{\eta_1}$.
Notice that, since we were considering the case $c_{13}=c_4=0$ in deriving Eqs.\eqref{master3a} and \eqref{master3b}, the corresponding $\eta_1$ and $\eta_2$ are happen to be identical under this condition.


\end{document}